\documentclass[11pt,eqs]{article}
\usepackage{latexsym,amsmath}
\usepackage{hyperref}
\usepackage{ulem}

\def\cleq{\setcounter{equation}{0}}
\title{Fluxes of Courant bracket twisted at the same time by $B$ and $\theta$ 
\thanks{Work supported in part by
the Serbian Ministry of Science, Technological Development and Innovation}}
\author{ Lj. Davidovi\'c \thanks{e-mail: ljubica@ipb.ac.rs}, I. Ivani\v sevi\'c \thanks{e-mail: ivanisevic@ipb.ac.rs} and B. Sazdovi\'c
\thanks{e-mail: sazdovic@ipb.ac.rs}\\
{\it Institute of Physics, University of Belgrade}\\
{\it Pregrevica 118, 11080 Belgrade, Serbia}}

\begin{document}
\maketitle
\begin{abstract}
This paper investigates the simultaneous twisting of the Courant bracket by a 2-form $B$ and a bi-vector $\theta$, exploring the generalized fluxes obtained in Courant algebroid relations. We define the twisted Lie bracket and demonstrate that the generalized $H$-flux can be expressed as the field strength defined on this Lie algebroid. Similarly, we show that the $f$-flux is a direct consequence of simultaneous twisting, which arises in the twisted Lie bracket relations between holonomic partial derivatives. We obtain the generalized $Q$ flux in terms of the twisted Koszul bracket, which is a quasi-Lie algebroid bracket. The action of an exterior derivative related to the twisted Koszul bracket on a bi-vector produces the generalized $R$-flux. We show that the generalized $R$-flux is also the twisted Schouten-Nijenhuis bracket, i.e. the natural graded bracket on multi-vectors defined with the twisted Lie bracket. 
\end{abstract}

\section{Introduction}
\cleq

The unification of vectors and 1-forms into generalized vectors has led to the development of generalized geometry, where various geometrical structures, including Poisson, symplectic, and Calabi-Yau manifolds, naturally emerge. The smooth section of the generalized tangent bundle is equipped with a Courant bracket, which reduces to the Lie bracket on pure vector fields \cite{courant}. In bosonic string theory, this skew-symmetric bracket emerges in the Poisson bracket algebra of a generator governing general coordinate and local gauge transformations \cite{doucou, cdual}. The two symmetries are generated by canonical variables - the former by coordinate $\sigma$-derivatives, while the latter by canonical momenta. The relation of these variables by T-duality allowed for interpretation of the Courant bracket as the extension of Lie bracket invariant under T-duality \cite{cdual}. 

The dynamics of a bosonic string propagating in spacetime fields, specifically the metric and the Kalb–Ramond field, is governed by a Hamiltonian that can be expressed using a generalized metric, which can subsequently be diagonalized via the so-called $B$-transformation. A similar approach applies to the T-dual theory, where the $\theta$-transformation reveals the diagonal T-dual generalized metric. These transformations, collectively referred to as twists, intermix the fundamental phase space variables, resulting in a basis formed by currents associated with the components of the energy-momentum tensors of the original and T-dual theories. The Poisson algebra structural constants of these currents encode the spacetime fluxes \cite{c, crdual}. 

String fluxes \cite{flux1} are important since they can create a potential that stabilizes the vacuum expectation value and provides mass to the massless non-physical particles called moduli \cite{flux-vac, flux-vac1, flux-vac2}. They describe geometry of different backgrounds connected by T-duality \cite{stw, dasgupta, danijel1, danijel2}. Moreover, fluxes are closely linked to the non-commutativity and non-associativity of string coordinates \cite{NC-string, NA-string}. The generalized fluxes were also obtained from generalized vielbeins \cite{flux2}, and were shown to measure the failure of twisted Courant brackets to satisfy Jacobi identity \cite{flux3}. 

Some of the well studied examples of fluxes from generalized geometry include the $B$-twisted Courant bracket, in which $H$-flux appears, and the $\theta$-twisted Courant bracket, in which $Q$ and $R$-fluxes appear. The twisting of the Courant bracket in general comes with the cost of losing the invariance of that bracket under T-duality. For instance, the $B$-twisted Courant bracket becomes the $\theta$-twisted Courant bracket when T-duality is performed \cite{crdual}. Moreover, there were some successful attempts to describe the topological T-duality as an isomorphism between Courant algebroids with different twisted Courant brackets as their underlining brackets \cite{Tdualiso}. Additionally, the T-duality between $H$ and $R$ fluxes via Courant algebroids was also described in \cite{HR-dual}. 

Since these twists are rooted in the properties of the fundamental T-duality rules, they motivate the consideration of a generalized twist, applied simultaneously using the Kalb–Ramond field and its T-dual counterpart, i.e. the non-commutativity parameter $\theta$. This should be the natural next step to multiple instances when the Courant bracket was firstly twisted by $B$, and then by $\theta$, resulting in the so called Roytenberg bracket, containing all generalized fluxes \cite{royt, royt1}. However, this bracket is not invariant under the T-duality, which is a consequence of the fact that $B$-shifts and $\theta$-transformations do not commute. That is why in \cite{CBTh}, the Courant bracket that was simultaneously twisted by $B$ and $\theta$ was constructed. The idea was that instead of twisting firstly by $B$ and then by $\theta$, simultaneous twisting would a priori define currents that mutually transform one into another under T-duality. In this paper, our goal is to obtain and analyse all the fluxes appearing in the algebra relations of the Courant bracket simultaneously twisted by both $B$ and $\theta$.

We begin the paper by outlining the basic mathematical notions of Lie algebroids, including the bracket, exterior derivative and graded generalization of Lie bracket on multi-vectors. We will introduce the Lie bracket generalization on the space of differential forms, called the Koszul bracket, as well as the Lie bracket generalization on the space of multi-vectors, called the Schouten-Nijenhuis bracket. These brackets can easily be generalized to arbitrary Lie algebroids, which will turn out to be useful for understanding the nature of the fluxes in the rest of the paper. Additionally, we will define the Courant bracket together with its twisted versions, and show the most well known examples of the Courant brackets twisted by $B$ and $\theta$, which are performed by $e^{\hat{B}}$ and $e^{\hat{\theta}}$ transformations, respectively.

In the third chapter, we will present the twist defining the Courant bracket simultaneously twisted by both $B$ and $\theta$. The key in construction is considering the matrix $\breve{B}$, which is a direct sum of matrices $\hat{B}$ and $\hat{\theta}$. The resulting twisting matrix features hyperbolic functions of the product of fields $\theta$ and $B$ and is very complicated transformation to work with. That is why we will introduce another $O(D, D)$ rotation, so that the currents have more recognizable form and are easier to work with. 

The expressions for generalized fluxes will be derived in the fourth chapter from the Poisson bracket algebra of the mutually T-dual currents. It will be demonstrated that the obtained generalized fluxes exhibit same relations as the fluxes obtained in the algebra relations of Roytenberg bracket.

In the fifth chapter, we will analyse geometries suitable for writing fluxes in the coordinate free notation. We introduce the twisted Lie bracket, as a bracket of the Lie algebroid defined with the hyperbolic function ${\cal C}$ as its anchor, and show that its algebra between holonomic partial derivatives gives rise to the $\breve{f}$-flux appearing in the algebra relations of currents. As for the $\breve{H}$-flux, we will express it as the action of the exterior derivative related to the twisted Lie bracket on the 2-form $\hat{B}$. 

In the following chapter, we also define the twisted Koszul bracket. We demand that the relation between the twisted Koszul bracket and the twisted Lie bracket mirrors the relation between their untwisted counterparts. This is a bracket of the twisted Lie algebroid. We demonstrate that the twisted Koszul bracket between two differential one-forms produces the $\breve{Q}$-flux. 

In addition, we introduce the twisted Schouten-Nijenhuis bracket as the graded extension of the twisted Lie bracket. We show that the twisted Schouten-Nijenhuis bracket of the appropriate bi-vector with itself produces exactly the $\breve{R}$-flux. Lastly, we demonstrate that these structures are necessary for the coordinate-free expression of the Courant brackets simultaneously twisted by $B$ and $\theta$.

\section{Mathematical preliminaries: from Lie to Courant algebroids}
\cleq

This mathematical introduction we begin with the notion of Lie algebroid. Broadly speaking, a Lie algebroid can be thought of as a structure in which the traditional tangent bundle is substituted with a vector bundle that shares similar properties. As a result, numerous geometrical concepts that traditionally rely on the tangent bundle are easily generalized to the context of Lie algebroids \cite{lie-alg-geometry}. These geometrical concepts will appear to be convenient to describe the fluxes that we are considering in this paper.

\subsection{Lie algebroid}

Lie algebroid \cite{lie-alg, lie-alg1} is a triple $\Big(V, [, ], \rho \Big)$ consisting of a vector bundle $V$, the anchor $\rho: V \to T{\cal M}$, and the skew-symmetric bracket $[,]$ on the space of smooth section of $V$, so that the following compatibility conditions are satisfied:
\begin{eqnarray}
&&\rho [\xi_1, \xi_2] = [\rho(\xi_1), \rho(\xi_2)]_L \, , \label{eq:lialg-def1}\\
&&[\xi_1, f \xi_2] = f [\xi_1, \xi_2] + ({\cal L}_{\rho(\xi_1)} f) \xi_2 \, ,\label{eq:lialg-def2} \\
&&[\xi_1, [\xi_2, \xi_3]] + [\xi_2, [\xi_3, \xi_1]] +[\xi_3, [\xi_1, \xi_2]]=0 \, , \label{eq:lialg-def3}
\end{eqnarray}
where $[ , ]_L$ denotes the usual Lie bracket. To gain some intuition, it is worth considering the most simple example of the Lie algebroid, consisting of the tangent bundle, the identity operator as the anchor $\rho = \text{Id}$ and the usual Lie bracket. The first condition (\ref{eq:lialg-def1}) is trivial, while the remaining conditions (\ref{eq:lialg-def2}) and (\ref{eq:lialg-def3}) are respectively well known Leibniz rule and Jacobi identity for the Lie bracket.

From the compatibility conditions in the Lie algebroid definition, we see that on this structure we can define a Lie derivative. Its action on functions is defined by
\begin{equation} \label{eq:calLf}
{\cal \hat{L}}_{\xi} f = {\cal L}_{\rho(\xi)} f \, ,
\end{equation}
and on vectors by the Lie algebroid bracket
\begin{equation} \label{eq:calLxi}
{\cal \hat{L}}_{\xi_1} \xi_2 = [\xi_1, \xi_2] \, .
\end{equation}

Additionally, we can introduce the analog of the exterior derivative. Recall that the exterior derivative extends the concept of the differential of a function to differential forms. The analog of the exterior derivative is a nilpotent operator given by \cite{lie-alg-geometry}
\begin{eqnarray} \label{eq:ext-der-rho}
\hat{d} \lambda (\xi_0, ... ,\xi_p) &=& \sum_{i=0}^p (-1)^{i} {\cal L}_{\rho(\xi_i)} \Big( \lambda(\xi_0, ... ,\hat{\xi}_i, ... ,\xi_p) \Big) \\ \notag
&&+ \sum_{i<j} (-1)^{i+j} \lambda([\xi_i, \xi_j], \xi_0, ... ,\hat{\xi}_i, ... ,\hat{\xi}_j, ..., \xi_p) \, .
\end{eqnarray}
Here, $\lambda$ is a $p$-covariant tensor from the section of dual space to the Lie algebroid bundle $V$. By $\lambda(\xi_0, ... ,\hat{\xi}_i, ... ,\xi_p) $ we marked this tensor contracted with $p$ different vectors $\xi_{0}\, , ... , \xi_{p}$, while the hat above the vector $\hat{\xi}_i$ denotes omission of that specific element. In case of a trivial Lie algebroid with identity anchor, the above relation defines the action of usual exterior derivative on $p$-forms. However, we should stress out that in case of general vector bundles, $\lambda$ should not be understood as $p$-forms. For example, if a Lie algebroid is defined on cotangent bundle, $\xi$ will be one-forms, while $\lambda$ will be $p$-vectors.

\subsubsection{Schouten-Nijenhuis bracket}

The Schouten-Nijenhuis bracket \cite{SNB} is a natural way to extend the generalization of the Lie bracket to the multi-vectors. It is defined by
\begin{equation} \label{eq:SNB-def}
[f, g]_S = 0 \, ,\quad [\xi, f]_S = {\cal L}_{\xi}(f) \, , \quad [\xi_1, \xi_2]_S = [\xi_1, \xi_2]_L \, ,
\end{equation}
where other brackets are obtained with the help of the graded Leibniz rule
\begin{equation} \label{eq:SNB-grad1}
[\theta_1, \theta_2 \wedge \theta_3]_S = [\theta_1, \theta_2 ]_S \wedge \theta_3 + (-1)^{(p-1)q}\theta_2 \wedge [\theta_1, \theta_3]_S \, ,  
\end{equation}
and the graded commutative relation
\begin{equation}\label{eq:SNB-grad2}
[\theta_1, \theta_2]_S = -(-1)^{(p-1)(q-1)}  [\theta_2, \theta_1]_S \, ,
\end{equation}
where $p$, and $q$ are orders of multi-vectors $\theta_1$ and $\theta_2$, respectively. 

It is also possible to obtain the coordinate expression for the Schouten-Nijenhuis bracket. Let 
\begin{eqnarray}
\theta_1 = \frac{1}{p!} \theta_1^{\mu_1, ... \mu_p} \partial_{\mu_1} \wedge ... \wedge \partial_{\mu_p} \, , \quad \theta_2 = \frac{1}{q!} \theta_2^{\nu_1, ... \nu_q} \partial_{\nu_1} \wedge ... \wedge \partial_{\nu_q} \, , 
\end{eqnarray}
then we have \cite{SNB}
\begin{eqnarray} \label{eq:SNB-coord}
[\theta_1, \theta_2]_{S} &=& \frac{1}{(p+q-1)!} [\theta_1, \theta_2]^{\mu_1 ... \mu_{p+q-1}}_{S} \partial_{\mu_1} \wedge ... \wedge \partial_{\mu_{p+q-1}} \, , \\ \notag
[\theta_1, \theta_2]^{\mu_1 ... \mu_{p+q-1}}_{S}  &=&\frac{1}{(p-1)!q!} \epsilon^{\mu_1 ... \mu_{p+q-1}}_{\nu_1 ... \nu_{p-1}\rho_1... \rho_{q}}\theta_1^{\sigma \nu_1 ... \nu_{p-1}}\partial_\sigma \theta_2^{\rho_1 ... \rho_q} \\ \notag
&&+\frac{(-1)^p}{p! (q-1)!} \epsilon^{\mu_1 ... \mu_{p+q-1}}_{\nu_1 ... \nu_{p}\rho_1... \rho_{q-1}}\theta_2^{\sigma \rho_1 ... \rho_{q-1}}\partial_\sigma \theta_1^{\nu_1 ... \nu_{p} } \, , 
\end{eqnarray}
where the antisymmetric Levi Civita symbol is defined by
\begin{eqnarray} \label{eq:levi-civita}
\epsilon^{\mu_1... \mu_p}_{\nu_1... \nu_p} = \begin{vmatrix}
\delta^{\mu_1}_{\nu_1} & ...& \delta^{\mu_1}_{\nu_p} \\
. & & .\\
. & & . \\
. & & . \\
\delta^{\mu_p}_{\nu_1} & ...& \delta^{\mu_p}_{\nu_p}
\end{vmatrix} \, .
\end{eqnarray}

The Lie algebroid generalization of the Schouten-Nijenhuis bracket is defined by
\begin{equation} \label{eq:SNB-liealg}
[f, g]_{{\bf S}} = 0 \, , \quad [\xi, f]_{{\bf S}} = {\cal L}_{\rho(\xi)}(f) \, , \quad [\xi_1, \xi_2]_{{\bf S}} = [\xi_1, \xi_2] \, ,
\end{equation}
with other brackets satisfying the graded identities (\ref{eq:SNB-grad1}) and (\ref{eq:SNB-grad2}). This is a bracket on a smooth section of higher product of vector bundle on which the Lie algebroid is defined. In case when the Lie algebroid bracket is defined on the smooth section of tangent bundle, its corresponding Schouten-Nijenhuis bracket will be defined on multi-vectors, but generally this is not the case. As such, the expression (\ref{eq:SNB-coord}) cannot be easily generalized for an arbitrary Lie algebroid, rather it has to be derived in each different case. 

\subsubsection{Koszul bracket}

Let us shortly illustrate the notion of Lie algebroid with a less trivial but typical (and for our paper important) example, consisting of the cotangent bundle $T^{\star}{\cal M}$ as its vector bundle. Elements of its smooth section are differential 1-forms $\lambda$. Let $\theta$ be a Poisson bi-vector, i.e. let $[\theta, \theta]_S = 0$, where
\begin{equation} \label{eq:theta-poisson}
\left. [\theta, \theta]_S \right| ^{\mu \nu \rho} = 2 (\theta^{\mu \sigma}\partial_{\sigma}\theta^{\nu \rho} + \theta^{\nu \sigma}\partial_{\sigma}\theta^{\rho \mu} +\theta^{\rho \sigma}\partial_{\sigma}\theta^{\mu \nu }) = 0 \, .
\end{equation}
This bi-vector defines a morphism by 
\begin{equation}
\theta(\lambda_1) \lambda_2 = \theta(\lambda_1, \lambda_2) \, , \quad \Big( \theta(\lambda_1)\Big)^{\! \mu} = \lambda_{1\nu} \theta^{\nu \mu} \, ,
\end{equation}
that satisfies 
\begin{equation}
\Big(\theta ([\lambda_1, \lambda_2]_{\theta}) \Big) = [\theta(\lambda_1), \theta(\lambda_2)]_L \, ,
\end{equation}
where we introduced the Koszul bracket \cite{koszul} which in the non-coordinate notation is given by
\begin{equation} \label{eq:koszul-def}
[\lambda_1, \lambda_2]_{\theta} = {\cal L}_{\theta (\lambda_1)}\lambda_2-{\cal L}_{\theta (\lambda_2)}\lambda_1-d (\theta(\lambda_1, \lambda_2)) \, ,
\end{equation}
or in some local basis by
\begin{equation} \label{eq:koszul-coord}
\Big( [\lambda_1, \lambda_2]_{\theta} \Big)_{\!\! \mu} = \theta^{\nu \rho} (\lambda_{1 \nu}\partial_\rho \lambda_{2 \mu}-\lambda_{2 \nu} \partial_\rho \lambda_{1 \mu})+ \lambda_{1 \rho} \lambda_{2 \nu} \partial_\mu \theta^{\rho \nu} \, .
\end{equation}
The cotangent bundle, with Koszul bracket and Poisson bi-vector as an anchor defines a Lie algebroid. The corresponding exterior derivative (\ref{eq:ext-der-rho}) acts on multi-vector. Its action can be expressed in terms of Schouten-Nijenhuis bracket by
\begin{equation} \label{eq:d-theta-s}
d_{\theta} = [\theta, . ]_{S} \, .
\end{equation}
Its square is zero, if and only if the bi-vector is Poisson bi-vector. 

Even in the case when a bi-vector is not Poisson bi-vector, i.e. the condition (\ref{eq:theta-poisson}) does not hold, it is still possible to define the Koszul bracket, even the exterior derivative of the Koszul bracket. However, in this instance, the Koszul bracket will not satisfy the Jacobi identity, and its corresponding exterior derivative will not be nilpotent. In this instance, we say that we have a quasi-Lie algebroid.

\subsection{Generalized tangent bundle}

The introduction of Lie algebroid allowed generalization of the geometric terms to arbitrary vector bundles. This can be generalized further by considering a direct sum of the tangent and cotangent bundle over a manifold $T{\cal M} \oplus T^{\star}{\cal M}$, which is called the generalized tangent bundle \cite{calabi-yau, gualtieri}. The elements of its section are generalized vectors 
\begin{equation}
\Lambda^M = \xi^{\mu} \oplus \lambda_{\mu} = \begin{pmatrix}
\xi^{\mu} \\
\lambda_{\mu}
\end{pmatrix} \, ,
\end{equation}
where $\xi$ represents its vector components, and $\lambda$ represents its 1-form components. The interior product defines a natural way to combine vectors and 1-forms into a scalar $i_{\xi} \lambda = \xi^\mu \lambda_\mu$. We can use this to define an inner product on the smooth section of the generalized tangent bundle by
\begin{equation} \label{eq:inner-pr}
\langle \Lambda_1, \Lambda_2 \rangle = \langle \xi_1 \oplus \lambda_1, \xi_2 \oplus \lambda_2 \rangle = i_{\xi_1} \lambda_2 + i_{\xi_2} \lambda_1 \, .
\end{equation}
The group of transformations that keeps this inner product invariant is $O(D, D)$ group \cite{ODD1, ODD2}, which is closely related to the string theory T-duality.

There is a natural extension of the Lie bracket on the smooth section of generalized tangent bundle called Courant bracket. Its expression is given by
\begin{eqnarray} \label{eq:courant-def}
[ \Lambda_1, \Lambda_2]_{\cal C} &=& \xi \oplus \lambda \\ \notag
\xi &=& [\xi_1, \xi_2]_L \, , \\ \notag
\lambda &=& {\cal L}_{\xi_1}\lambda_2- {\cal L}_{\xi_2}\lambda_1 - \frac{1}{2} d (i_{\xi_1}\lambda_2-i_{\xi_2}\lambda_1) \, ,
\end{eqnarray}
or
\begin{eqnarray} \label{eq:courant-coord}
\xi^\mu &=& \xi_1^\nu \partial_\nu \xi_2^\mu - \xi_2^\nu \partial_\nu \xi_1^\mu \, , \\ \notag
\lambda_\mu &=& \xi_1^\nu (\partial_\nu \lambda_{2 \mu} - \partial_\mu \lambda_{2 \nu}) - \xi_2^\nu (\partial_\nu \lambda_{1 \mu} - \partial_\mu \lambda_{1 \nu})+\frac{1}{2} \partial_\mu (\xi_1 \lambda_2- \xi_2 \lambda_1 ) \, .
\end{eqnarray}
This bracket is known to appear in string theory context in the Poisson bracket algebra of symmetry generators. Symmetries in string theory $\sigma$-model are governed by generators that in classical theory act on Hamiltonian via Poisson bracket. General coordinate transformations, or diffeomorphisms, are generated by
\cite{dualsim}
\begin{equation} \label{eq:GCTdef}
{\cal G}_{\xi} =\int_0^{2\pi} d \sigma \xi^\mu (x(\sigma)) \pi_\mu (\sigma) \, ,
\end{equation}
that leads to string background field transformations
\begin{eqnarray} \label{eq:delta-xi}
\delta_{\xi} G_{\mu \nu} &=& {\cal L}_{\xi} G_{\mu \nu} = \xi^\rho \partial_\rho G_{\mu \nu} + \partial_\mu \xi^\rho G_{\rho \nu} +\partial_\nu \xi^\rho G_{\rho \mu} \, ,\\ 
\delta_{\xi} B_{\mu \nu} &=& {\cal L}_{\xi} B_{\mu \nu} = \xi^\rho \partial_\rho B_{\mu \nu} - \partial_\mu \xi^\rho B_{\rho \nu}+ B_{\mu \rho}\partial_\nu \xi^\rho \, .
\end{eqnarray}
Similarly, local gauge transformations are generated by \cite{dualsim}
\begin{equation}\label{eq:GenLG}
{\cal G}_{\lambda} =\int_0^{2\pi} d \sigma  \lambda_\mu(x(\sigma)) \kappa x^{\prime\mu}(\sigma) \, ,
\end{equation}
which results in
\begin{eqnarray} \label{eq:delta-lambda}
\delta_{\lambda} G_{\mu \nu} &=& 0 \, ,\\ \notag
\delta_{\lambda} B_{\mu \nu} &=& (d \lambda)_{\mu \nu} = \partial_\mu \lambda_\nu - \partial_\nu \lambda_\mu \, .
\end{eqnarray}
Taking the joint generator of diffeomorphisms and local gauge transformations, we obtain
\begin{equation} \label{eq:Gen-def}
{\cal G}_{\Lambda} = {\cal G}_{\xi} +{\cal G}_{\lambda} = \int d\sigma \langle \Lambda, X \rangle \, , 
\end{equation}
where
\begin{equation}
X^M = \begin{pmatrix}
\kappa x^{\prime \mu} \\
\pi_\mu
\end{pmatrix} \, , 
\end{equation}
and
\begin{equation}
\Lambda^M = \begin{pmatrix}
\xi^{\mu} \\
\lambda_\mu
\end{pmatrix} \, .
\end{equation}
The Courant bracket appears in the Poisson bracket relation of this joint generator
\begin{equation}
\Big\{ {\cal G}_{\Lambda_1}, \, {\cal G}_{\Lambda_2} \Big\} = -{\cal G}_{[\Lambda_1, \Lambda_2]_{\cal C}}\, .
\end{equation}
%FALE REFERENCE!!1

\subsection{Twisted Courant bracket}

The Courant bracket is not the only choice of bracket on generalized tangent bundle. Various twists of the bracket can be obtained, by which different string fluxes appear. The simplest example is the $B$-twisted Courant bracket
\begin{equation} \label{eq:B-bracket-def}
[\Lambda_1, \Lambda_2]_{{\cal C}_B} = e^{\hat{B}}[e^{-\hat{B}}\Lambda_1, e^{-\hat{B}}\Lambda_2]_{\cal C} \, ,
\end{equation}
where $e^{\hat{B}}$ denotes $B$-transformation, given by
\begin{equation} \label{eq:ebhat}
e^{\hat{B}} = \begin{pmatrix}
\delta^\mu_\nu & 0 \\
2B_{\mu \nu} & \delta^\nu_\mu
\end{pmatrix}\, , \ \ 
\hat{B}^M_{\ N} = 
\begin{pmatrix}
0 & 0 \\
2B_{\mu \nu} & 0 \\ 
\end{pmatrix}\, .
\end{equation}
This bracket can be obtained by expressing the generator (\ref{eq:Gen-def}) in the basis 
\begin{equation} \label{eq:XhatB}
\hat{X}^M = (e^{\hat{B}})^M_{\ N}\ X^N = 
\begin{pmatrix}
\kappa x^{\prime \mu} \\
i_\mu
\end{pmatrix}
\, ,
\end{equation}
where $i_\mu$ are currents given by
\begin{equation}\label{eq:idef}
i_{\mu} = \pi_\mu + 2\kappa B_{\mu \nu}x^{\prime \nu} \, .
\end{equation}
The Poisson bracket between these currents gives rise to the relation \cite{c, cdual}
\begin{equation} \label{eq:iialgebra}
\{ i_\mu (\sigma), i_\nu (\bar{\sigma}) \} = - 2\kappa B_{\mu \nu \rho} x^{\prime \rho} \delta (\sigma - \bar{\sigma}) \, ,
\end{equation}
where $B_{\mu \nu \rho}$ is the Kalb-Ramond field strength, or H-flux, given by
\begin{equation}
B_{\mu \nu \rho}  = \partial_\mu B_{\nu \rho} + \partial_\nu B_{\rho \mu} +\partial_\rho B_{\mu \nu} \, . 
\end{equation}

Another example is the $\theta$-twisted Courant bracket 
\begin{equation} \label{eq:th-bracket-def}
[\hat{\Lambda}_1, \hat{\Lambda}_2]_{{\cal C}_{\theta}} = e^{\hat{\theta}}[e^{-\hat{\theta}}\Lambda_1, e^{-\hat{\theta}}\Lambda_2]_{\cal C} \, ,
\end{equation}
where twisting is performed by the $\theta$-transformation, given by
\begin{equation} \label{eq:ehatth}
e^{\hat{\theta}} = \begin{pmatrix}
\delta^\mu_\nu & \kappa \theta^{\mu \nu} \\
0 & \delta^\nu_\mu 
\end{pmatrix} \, , \quad \hat{\theta}^M_{\ N} = \begin{pmatrix}
0 & \kappa \theta^{\mu \nu} \\
0 & 0 
\end{pmatrix} \, .
\end{equation}
The Courant bracket twisted by $\theta$ can be obtained if the generator  (\ref{eq:Gen-def}) is expressed in basis defined by
\begin{equation}
\hat{X}^M = (e^{\hat{\theta}})^M_{\ N}\ X^N = 
\begin{pmatrix}
k^\mu \\
\pi_\mu 
\end{pmatrix} \, ,
\end{equation}
where $k^\mu$ are currents, given by \cite{cdual}
\begin{eqnarray}\label{eq:defk}
k^\mu = \kappa  x^{\prime\mu}    + \kappa  \theta^{\mu\nu} \pi_\nu   \,  .
\end{eqnarray}
The algebra of these currents closes on a different set of fluxes, i.e.
\begin{eqnarray} \label{eq:pbrk}
\{ k^\mu (\sigma), k^{\nu} (\bar{\sigma}) \} =-\kappa Q_\rho^{\ \mu \nu} k^\rho \delta(\sigma - \bar{\sigma}) - \kappa^2 R^{\mu \nu \rho} \pi_\rho \delta (\sigma - \bar{\sigma}) \, ,
\end{eqnarray}
where $Q$ and $R$ fluxes are given by
\begin{equation} \label{eq:QRdef}
Q_{\rho}^{\ \mu \nu} = \partial_\rho \theta^{\mu \nu}, \ \ \ \ \ R^{\mu \nu \rho} = \theta^{\mu \sigma} \partial_\sigma \theta^{\nu \rho} + \theta^{\nu \sigma} \partial_\sigma \theta^{\rho \mu} + \theta^{\rho \sigma} \partial_\sigma \theta^{\mu \nu} \, .
\end{equation}
In addition, the Courant bracket can be twisted by other more general multi-vectors \cite{keremcan}.

Interesting insights emerge when one takes into account T-duality, under which momenta and winding around a compact dimension are swapped. The former are integrals of canonical momenta $\pi_\mu$, while the latter are integrals of the coordinate $\sigma$-derivatives $\kappa x^{\prime \mu}$ around the compact dimension. Moreover, the Buscher T-dualization rules \cite{buscher, buscher1} define the T-dual of the Kalb-Ramond field $B$ to be the non-commutativity parameter $\theta$. These T-duality relations we write as
\begin{equation} \label{eq:t-dual}
\kappa x^{\prime \mu} \cong \pi_\mu \, , \qquad B_{\mu \nu} \cong  \frac{\kappa}{2} \theta^{\mu \nu} \, .
\end{equation}
When these interchanges are made, the currents $i_\mu$ (\ref{eq:idef}) transform into currents $k^\mu$ (\ref{eq:defk}), i.e.
\begin{equation} \label{eq:ik-dual}
i_\mu \cong k^\mu \, .
\end{equation}
Given the relations (\ref{eq:t-dual}) and (\ref{eq:ik-dual}), we can conclude that the $B$-twisted Courant and $\theta$-twisted Courant bracket are in fact related by the T-duality \cite{crdual}. 

The T-dual relations between these twisted Courant brackets can also be obtained from the double theory. In that instance, the generalization of the generator (\ref{eq:Gen-def}) gives rise to the $C$-bracket \cite{siegel, siegel1}, which by the same method can be twisted, giving rise to the twisted $C$-brackets \cite{TwistedC}. When these twisted $C$-brackets are projected to mutually T-dual phase spaces, the $B$-twisted and $\theta$-twisted Courant brackets are respectively obtained.

\section{Courant bracket twisted simultaneously by both B and $\theta$}
\cleq

The Courant bracket twisted simultaneously by both B and $\theta$ was constructed in \cite{CBTh} with the goal of obtaining the twisted Courant bracket containing all fluxes, while simultaneously being invariant under the T-duality. In this chapter, we will outline the overview of its construction. For details, reader is referred to the reference \cite{CBTh}. 

The twisting matrix is defined in the form $e^{\breve{B}}$, where
\begin{equation} \label{eq:breve}
\breve{B} = \hat{B}+\hat{\theta} = 
\begin{pmatrix}
0 & \kappa \theta^{\mu \nu} \\
2 B_{\mu \nu} & 0
\end{pmatrix} \, .
\end{equation}
Given that the Kalb-Ramond field $B$ is T-dual to the non-commutativity parameter $\theta$, this transformation is invariant under T-duality by design. The expression for $e^{\breve{B}}$ is obtained from its Taylor expansion
\begin{equation} \label{eq:ebb}
e^{\breve{B}} = 
\begin{pmatrix}
{\cal C}^{\mu}_{\ \nu} & \kappa {\cal S}^\mu_{\ \rho} \theta^{\rho \nu} \\
2 B_{\mu \rho} {\cal S}^{\rho}_{\ \nu} & ( {\cal C}^T)^{\ \nu}_{\mu}
\end{pmatrix} \, ,
\end{equation}
where
\begin{equation} \label{eq:CS-def}
{\cal C}^\mu_{\ \nu} = \cosh{\sqrt{\alpha}}^\mu_{\ \nu} \, , \quad {\cal S}^\mu_{\ \nu} = \Big(\frac{\sinh{\sqrt{\alpha}}}{\sqrt{\alpha}} \Big)^\mu_{\ \nu} \, ,
\end{equation}
and
\begin{equation}
\alpha^\mu_{\ \nu} = 2\kappa \theta^{\mu \rho} B_{\rho \nu} \, .
\end{equation}
Under T-duality we have $\alpha \cong \alpha^T$, which easily generalizes to all analytic functions of $\alpha$
\begin{equation} \label{eq:CSdual}
{\cal C} \cong {\cal C}^T \, , \ \ {\cal S} \cong {\cal S}^T \, .
\end{equation} 

The Courant bracket twisted at the same time by $B$ and $\theta$ can be obtained from the Poisson bracket algebra of the generator
\begin{equation} \label{eq:breveG}
{\cal G}_{\breve{\Lambda}} = \int d\sigma \langle \breve{\Lambda}, \breve{X} \rangle \, ,
\end{equation}
which is expressed in the basis 
\begin{equation} \label{eq:breveX}
\breve{X}^M = (e^{\breve{B}})^{\!M}_{\ N}\ X^N = \begin{pmatrix}
\breve{k}^\mu \\
\breve{\iota}_\mu
\end{pmatrix} \, ,
\end{equation}
where new currents are
\begin{eqnarray} \label{eq:i-k-breve}
\breve{k}^\mu &=& \kappa {\cal C}^\mu_{\ \nu} x^{\prime \nu} + \kappa ({\cal S} \theta)^{\mu \nu} \pi_\nu \, , \\ \notag
\breve{\iota}_\mu &=& 2  (B{\cal S})_{\mu \nu} x^{\prime \nu} + ( {\cal C}^T)^{\ \nu}_{\mu} \pi_\nu \, .
\end{eqnarray}
Due to (\ref{eq:CSdual}), these currents are mutually related by T-duality.

It is convenient to introduce an auxiliary basis defined by
\begin{equation} \label{eq:ik-ik}
\mathring{k}^\mu = {\cal C}^\mu_{\ \nu} \breve{k}^\nu = \kappa x^{\prime \mu} + \kappa \mathring{\theta}^{\mu \nu} \mathring{\iota}_\nu\, , \quad \mathring{\iota}_\mu = \breve{\iota}_\nu ({\cal C}^{-1})^\nu_{\ \mu} =  \pi_\mu + 2\kappa \mathring{B}_{\mu \nu} x^{\prime \nu} \, ,
\end{equation}
where
\begin{equation}
\mathring{B}_{\mu \nu} = (B{\cal S} {\cal C}^{-1})_{\mu \nu} \, , \quad \mathring{\theta}^{\mu \nu} = ( {\cal C}{\cal S} \theta)^{\mu \nu} \, .
\end{equation}
The currents $\mathring{k}^\mu$ and $\mathring{\iota}_\mu$ have the same form as the expressions that are often investigated when twisted Courant brackets with all fluxes are considered \cite{flux2, flux3, nick1, royt, royt1}. The relevant Poisson bracket relations between auxiliary currents are given by \cite{CBTh}
\begin{eqnarray} \label{eq:iotaiota}
&& \{ \mathring{\iota}_\mu (\sigma), \mathring{\iota}_\nu (\bar{\sigma})\} = - 2{\cal \mathring{B}}_{\mu \nu \rho} \mathring{k}^{\rho} \delta(\sigma-\bar{\sigma})- {\cal \mathring{F}}_{\mu \nu}^{\ \rho}\   \mathring{\iota}_\rho \delta(\sigma-\bar{\sigma}) \, , \\ \label{eq:kringkring}
&& \{ \mathring{k}^\mu (\sigma), \mathring{k}^\nu (\bar{\sigma}) \} = - \kappa{\cal \mathring{Q}}_{\rho}^{\ \mu \nu} \mathring{k}^\rho \delta(\sigma-\bar{\sigma}) -\kappa^2{\cal \mathring{R}}^{\mu \nu \rho} \mathring{\iota}_\rho \delta(\sigma-\bar{\sigma}) \, ,\\ \label{eq:iotakring}
&& \{ \mathring{\iota}_\mu (\sigma), \mathring{k}^\nu (\bar{\sigma}) \} = \kappa \delta_\mu^\nu \delta^\prime(\sigma-\bar{\sigma}) + {\cal \mathring{F}}^{\ \nu}_{\mu \rho}\ \mathring{k}^\rho \delta(\sigma-\bar{\sigma}) -\kappa {\cal \mathring{Q}}_{\mu}^{\ \nu \rho} \mathring{\iota}_\rho \delta(\sigma-\bar{\sigma}) \, ,
\end{eqnarray}
where the auxiliary generalized fluxes are given by 
\begin{eqnarray} \label{eq:fluxring}
{\cal \mathring{B}}_{\mu \nu \rho} &=& \partial_\mu \mathring{B}_{\nu \rho} + \partial_\nu \mathring{B}_{\rho \mu}+\partial_\rho \mathring{B}_{\mu \nu} \, , \\ \label{eq:calFring}
{\cal \mathring{F}}_{\mu \nu}^{\ \rho} &=& -2\kappa {\cal \mathring{B}}_{\mu \nu \sigma} \mathring{\theta}^{\sigma \rho} \, , \\ \label{eq:calQring}
{\cal \mathring{Q}}^{\ \nu \rho}_\mu
 &=&   \mathring{Q}^{\ \nu \rho}_\mu + 2\kappa \mathring{\theta}^{\nu\sigma} \mathring{\theta}^{\rho \tau} {\cal \mathring{B}}_{\mu \sigma \tau} \, , \quad  \mathring{Q}^{\ \nu \rho}_\mu = \partial_{\mu} \mathring{\theta}^{\nu \rho} \\ \label{eq:calRring}
{\cal \mathring{R}}^{\mu \nu \rho} &=& \mathring{R}^{\mu \nu \rho} +2\kappa \mathring{\theta}^{\mu \lambda} \mathring{\theta}^{\nu \sigma}\mathring{\theta}^{\rho \tau}{\cal \mathring{B}}_{\lambda \sigma \tau} \, , \\ \notag
 \mathring{R}^{\mu \nu \rho} &=& \mathring{\theta}^{\mu \sigma}\partial_\sigma \mathring{\theta}^{\nu \rho} + \mathring{\theta}^{\nu \sigma}\partial_\sigma \mathring{\theta}^{\rho \mu}+ \mathring{\theta}^{\rho \sigma}\partial_\sigma \mathring{\theta}^{\mu \nu}  \, .
\end{eqnarray}
The formulae (\ref{eq:fluxring})-(\ref{eq:calRring}) are the well known expressions for generalized fluxes. These fluxes, as well as their Bianchi identities, were obtained from the non-associative quasi-Poisson structures \cite{flux3}. 

\section{Generalized fluxes}

The algebra of auxiliary currents closes with auxiliary fluxes as its structure functions. We can now use the expression for currents $\breve{k}$ and $\breve{\iota}$ in terms of their auxiliary counterparts (\ref{eq:ik-ik}) to compute the fluxes relevant for the Courant bracket twisted with both $B$ and $\theta$. Besides the relations (\ref{eq:fluxring})-(\ref{eq:calRring}), we will require the algebra relations in the form 
\begin{equation} \label{eq:aux-C}
\{ {\cal C}^\mu_{\ \rho}(\sigma) , \mathring{\iota}_\nu (\bar{\sigma}) \} = \partial_\nu {\cal C}^\mu_{\ \rho} \delta(\sigma-\bar{\sigma}) \, , \quad \{ {\cal C}^\mu_{\ \rho}(\sigma) , \mathring{k}^\nu (\bar{\sigma}) \} = \kappa \mathring{\theta}^{\nu \sigma}\partial_\sigma {\cal C}^\mu_{\ \rho} \delta(\sigma-\bar{\sigma}) \, .
\end{equation}

Let us begin with writing the Poisson bracket relation between two sets of currents $\breve{\iota}_\mu$
\begin{eqnarray} \label{eq:iota-iota}
\{ \breve{\iota}_\mu (\sigma), \breve{\iota}_\nu (\bar{\sigma}) \} &=&  {\cal C}^\rho_{\ \mu} {\cal C}^\sigma_{\ \nu}   \{ \mathring{\iota}_\rho , \mathring{\iota}_\sigma \} +{\cal C}^\rho_{\ \mu}  \mathring{\iota}_\sigma   \{ \mathring{\iota}_\rho ,{\cal C}^\sigma_{\ \nu} \} + \mathring{\iota}_\rho  {\cal C}^\sigma_{\ \nu}  \{ {\cal C}^\rho_{\ \mu} , \mathring{\iota}_\sigma  \}  \\ \notag
&=& -2 {\cal C}^\rho_{\ \mu} {\cal C}^\sigma_{\ \nu}   \mathring{B}_{\rho \sigma \alpha} \mathring{k}^\alpha \delta(\sigma-\bar{\sigma}) - {\cal C}^\rho_{\ \mu} {\cal C}^\sigma_{\ \nu}   \cal{\mathring{F}}_{\!\!\rho \sigma}^{\ \alpha}\ \mathring{\iota}_\alpha \delta(\sigma-\bar{\sigma}) \\ \notag
&& - \Big( {\cal C}^\rho_{\ \mu}   \partial_\rho {\cal C}^\sigma_{\ \nu} -{\cal C}^\rho_{\ \nu}   \partial_\rho {\cal C}^\sigma_{\ \mu} \Big)\mathring{\iota}_\sigma \delta(\sigma-\bar{\sigma})  \\ \notag
&=& -2 \breve{\cal B}_{\mu \nu \rho} \breve{k}^{\rho} \delta(\sigma-\bar{\sigma}) - \breve{\cal F}_{\!\!\mu \nu}^{\ \rho}\ \breve{\iota}_\rho  \delta(\sigma-\bar{\sigma}) \, .
\end{eqnarray}
We firstly used (\ref{eq:ik-ik}), after which we substituted (\ref{eq:iotaiota}) and (\ref{eq:aux-C}). The resulting algebra we expressed in terms of $\breve{{\cal B}}$ flux, given by
\begin{equation} \label{eq:breveB}
\breve{\cal B}_{\mu \nu \rho} = {\cal C}^\alpha_{\ \mu}  {\cal C}^\beta_{\ \nu}  {\cal C}^\gamma_{\ \rho} \mathring{B}_{\alpha \beta \gamma} \, ,
\end{equation}
and $\breve{{\cal F}}$ flux, given by
\begin{equation} \label{eq:breveF}
\breve{\cal F}_{\!\!\mu \nu}^{\ \rho} = \mathring{\cal F}_{\!\!\alpha \beta}^{\ \gamma} {\cal C}^{\alpha}_{\ \mu} {\cal C}^{\beta}_{\ \nu} ({\cal C}^{-1})^\rho_{\ \gamma} + ({\cal C}^{-1})^{\rho}_{\ \tau} \Big( {\cal C}^{\sigma}_{\ \mu} \partial_\sigma {\cal C}^{\tau}_{\ \nu}-   {\cal C}^{\sigma}_{\ \nu} \partial_\sigma {\cal C}^{\tau}_{\ \mu}\Big) \, .
\end{equation}

In order to simplify the expression for $\breve{{\cal F}}$ flux, let us define
\begin{equation} \label{eq:hatpartial}
\hat{\partial}_\mu = ({\cal C}^T)_\mu^{\ \nu}\ \partial_\nu \, .
\end{equation}
Recall that $({\cal C}^T)_\mu^{\ \nu} $ is a hyperbolic cosine function (\ref{eq:CS-def}) of a matrix $\alpha$. It is non-zero, and there is a Taylor's expansion of its inverse that is defined everywhere. This means that we can invert the relation (\ref{eq:hatpartial}), where we obtain 
\begin{equation}
\partial_\mu =  ({\cal C}^{-T})_\mu^{\ \nu}\ \hat{\partial}_\nu \, ,
\end{equation}
where $({\cal C}^{-T}) = ({\cal C}^T)^{-1}$. We can rewrite the bi-vector $\mathring{\theta}$ in this new basis by
\begin{equation}
\mathring{\theta}^{\mu \nu} \partial_\mu \partial_\nu = \breve{\theta}^{\mu \nu} \hat{\partial}_\mu \hat{\partial}_\nu \, ,
\end{equation}
from which we obtain 
\begin{equation} \label{eq:brthdef}
\breve{\theta}^{\mu \nu} = ({\cal S C}^{-1})^\mu_{\ \rho} \theta^{ \rho\nu} = (C^{-2})^\mu_{\ \rho} \mathring{\theta}^{\rho \nu} \, .
\end{equation}
After substituting (\ref{eq:calFring}) and (\ref{eq:brthdef}) into (\ref{eq:breveF}), one obtains 
\begin{equation} \label{eq:breveFF}
\breve{\cal F}_{\!\!\mu \nu}^{\ \rho} = \breve{f}_{\!\mu \nu}^{\ \rho} - 2\kappa \breve{{\cal B}}_{\mu \nu \sigma} \breve{\theta}^{\sigma \rho} \, , 
\end{equation}
where
\begin{equation} \label{eq:brevef}
\breve{f}_{\!\mu \nu}^{\ \rho}  =  ({\cal C}^{-1})^{\rho}_{\ \sigma} \Big( \hat{\partial}_\mu{\cal C}^{\sigma}_{\ \nu}-   \hat{\partial}_\nu {\cal C}^{\sigma}_{\ \mu}\Big) \, .
\end{equation}

Now we can proceed with calculating the algebra of auxiliary currents $\breve{k}$. Starting with (\ref{eq:ik-ik}), with the help of (\ref{eq:kringkring}) and (\ref{eq:aux-C}), we have
\begin{eqnarray} \label{eq:kappa-kappa}
\{ \breve{k}^\mu(\sigma) , \breve{k}^\nu (\bar{\sigma})  \} &=& ({\cal C}^{-1})^\mu_{\ \rho} ({\cal C}^{-1})^\nu_{\ \sigma}  \{  \mathring{k}^\rho, \mathring{k}^\sigma  \} +({\cal C}^{-1})^\mu_{\ \rho} \mathring{k}^\sigma  \{  \mathring{k}^\rho ,({\cal C}^{-1})^\nu_{\ \sigma}  \} \\ \notag
&&+\mathring{k}^\rho \{ ({\cal C}^{-1})^\mu_{\ \rho}  ,\mathring{k}^\sigma \} ({\cal C}^{-1})^\nu_{\ \sigma}   \\ \notag
&=& -\kappa ({\cal C}^{-1})^\mu_{\ \rho} ({\cal C}^{-1})^\nu_{\ \sigma} \mathring{Q}_{\tau}^{\ \rho \sigma}\mathring{k}^\tau \delta(\sigma-\bar{\sigma})  - \kappa^2 ({\cal C}^{-1})^\mu_{\ \rho} ({\cal C}^{-1})^\nu_{\ \sigma} \mathring{R}^{\rho \sigma \tau} \mathring{\iota}_\tau \delta(\sigma-\bar{\sigma})  \\ \notag
&& - \kappa \Big( \breve{\theta}^{\mu \alpha} \hat{\partial}_\alpha ({\cal C}^{-1})^\nu_{\ \rho} -  \breve{\theta}^{\nu \alpha} \hat{\partial}_\alpha ({\cal C}^{-1})^\mu_{\ \rho} \Big) \mathring{k}^{\rho} \delta(\sigma-\bar{\sigma})  \\ \notag
&=& -\kappa \breve{\cal Q}_\rho^{\ \mu \nu} \breve{k}^\rho  \delta(\sigma-\bar{\sigma})  - \kappa^2 \breve{{\cal R}}^{\mu \nu \rho} \breve{\iota}_\rho  \delta(\sigma-\bar{\sigma}) \, ,
\end{eqnarray}
where we used the relations (\ref{eq:hatpartial}) and (\ref{eq:brthdef}) to simplify result. The fluxes that we obtained are
\begin{equation} \label{eq:breveQ}
\breve{\cal Q}_\rho^{\ \mu \nu} = {\cal C}^\alpha_{\ \rho} ({\cal C}^{-1})^\mu_{\ \beta} ({\cal C}^{-1})^\nu_{\ \gamma} \mathring{\cal Q}_\alpha^{\ \beta \gamma} - {\cal C}^{\alpha}_{\ \rho} \Big( \breve{\theta}^{\nu \beta} \hat{\partial}_\beta ({\cal C}^{-1})^\mu_{\ \alpha}- \breve{\theta}^{\mu \beta} \hat{\partial}_\beta ({\cal C}^{-1})^\nu_{\ \alpha} \Big)  \, ,
\end{equation}
and
\begin{equation} \label{eq:breveR}
\breve{\cal R}^{\mu \nu \rho} = \mathring{\cal R}^{\alpha \beta \gamma} ({\cal C}^{-1})^\mu_{\ \alpha}  ({\cal C}^{-1})^\nu_{\ \beta}  ({\cal C}^{-1})^\rho_{\ \gamma} \, .
\end{equation}
These are expressions for analogs of $Q$ and $R$ fluxes (\ref{eq:QRdef}). We will now proceed to rewrite them in more recognizable forms. For $ \breve{\cal{Q}}$-flux, substituting (\ref{eq:calQring}) into (\ref{eq:breveQ}), we obtain 
\begin{equation} \label{eq:breveQQ}
\breve{\cal Q}_\rho^{\ \mu \nu} = \breve{Q}_\rho^{\ \mu \nu} + 2\kappa \breve{\theta}^{\mu \alpha} \breve{\theta}^{\nu \beta} {\cal \breve{B}}_{\rho \alpha \beta } \, ,
\end{equation}
where
\begin{equation}
\breve{Q}_\rho^{\ \mu \nu} = {\cal C}^\alpha_{\ \rho} ({\cal C}^{-1})^\mu_{\ \beta} ({\cal C}^{-1})^\nu_{\ \gamma} \mathring{ Q}_\alpha^{\ \beta \gamma} - {\cal C}^{\alpha}_{\ \rho} \Big( \breve{\theta}^{\nu \beta} \hat{\partial}_\beta ({\cal C}^{-1})^\mu_{\ \alpha}- \breve{\theta}^{\mu \beta} \hat{\partial}_\beta ({\cal C}^{-1})^\nu_{\ \alpha} \Big) \, .
\end{equation}
After expressing the previous relation in terms of $\breve{\theta}$ (\ref{eq:brthdef}), we obtain
\begin{eqnarray} \label{eq:Qbreve}
\breve{Q}_\rho^{\ \mu \nu} &=& \partial_\alpha ({\cal C}^2 \breve{\theta})^{\beta \gamma} {\cal C}^\alpha_{\ \rho} ({\cal C}^{-1})^\mu_{\ \beta}({\cal C}^{-1})^\nu_{\ \gamma}-{\cal C}^\alpha_{\ \rho} \Big( \breve{\theta}^{\nu \beta} \hat{\partial}_\beta ({\cal C}^{-1})^\mu_{\ \alpha}- \breve{\theta}^{\mu \beta} \hat{\partial}_\beta ({\cal C}^{-1})^\nu_{\ \alpha} \Big) \\ \notag
&=& \hat{\partial}_\rho \breve{\theta}^{\mu \nu} - \Big( ({\cal C}\breve{\theta})^{\beta \nu} \hat{\partial}_\rho  ({\cal C}^{-1})^\mu_{\ \beta} + ({\cal C}\breve{\theta})^{\mu \gamma} \hat{\partial}_\rho ({\cal C}^{-1})^\nu_{\ \gamma} ) \Big)   \\ \notag 
&& -{\cal C}^\alpha_{\ \rho} \Big( \breve{\theta}^{\nu \beta} \hat{\partial}_\beta ({\cal C}^{-1})^\mu_{\ \alpha}- \breve{\theta}^{\mu \beta} \hat{\partial}_\beta ({\cal C}^{-1})^\nu_{\ \alpha} \Big)  \\ \notag
&=& \hat{\partial}_\rho \breve{\theta}^{\mu \nu}   + {\cal C}^\alpha_{\ \beta} {\breve{\theta}}^{\nu \beta} \hat{\partial}_\rho({\cal C}^{-1})^\mu_{\ \alpha}- {\cal C}^\alpha_{\ \beta}  {\breve{\theta}}^{\mu \beta} \hat{\partial}_\rho ({\cal C}^{-1})^\nu_{\ \alpha}\\ \notag 
&& -{\cal C}^\alpha_{\ \rho} \Big( \breve{\theta}^{\nu \beta} \hat{\partial}_\beta ({\cal C}^{-1})^\mu_{\ \alpha}- \breve{\theta}^{\mu \beta} \hat{\partial}_\beta ({\cal C}^{-1})^\nu_{\ \alpha} \Big)
\\ \notag
&=& \hat{\partial}_\rho \breve{\theta}^{\mu \nu}   + {\breve{\theta}}^{\nu \beta}\Big({\cal C}^\alpha_{\ \beta}  \hat{\partial}_\rho({\cal C}^{-1})^\mu_{\ \alpha}-{\cal C}^\alpha_{\ \rho}\hat{\partial}_\beta ({\cal C}^{-1})^\mu_{\ \alpha}\Big)\\ \notag 
&& - {\breve{\theta}}^{\mu \beta} \Big({\cal C}^\alpha_{\ \beta}  \hat{\partial}_\rho ({\cal C}^{-1})^\nu_{\ \alpha} -{\cal C}^\alpha_{\ \rho}\hat{\partial}_\beta ({\cal C}^{-1})^\nu_{\ \alpha} \Big)
\\ \notag
&=&\hat{\partial}_\rho \breve{\theta}^{\mu \nu} + \breve{f}^{\ \mu}_{\!\rho \sigma}\  \breve{\theta}^{\sigma \nu} - \breve{f}^{\ \nu}_{\!\rho \sigma}\ \breve{\theta}^{\sigma \mu} \, . 
\end{eqnarray}
Let us unpack what happened in the previous relation. In the first step, we expressed the flux $\mathring{Q}$ using the non-commutative field $\breve{\theta}$. Then, in the second step, we used partial integration on the first term and rearranged the resulting expression using equation (\ref{eq:hatpartial}). In subsequent steps, we recognized that $({\cal C}\breve{\theta})^{\beta \nu}$ can be expressed as ${\cal C}^\beta_{\ \sigma} \breve{\theta}^{\sigma \nu}$ and that ${\cal C}^\mu_{\ \beta} \partial_\alpha ({\cal C}^{-1})^\beta_{\ \sigma}$ equals $-\partial_\alpha {\cal C}^\mu_{\ \beta} ({\cal C}^{-1})^\beta_{\ \sigma}$. By relabeling some indices and using equation (\ref{eq:brevef}), we arrived at the final step of equation (\ref{eq:Qbreve}).

Similarly, substituting (\ref{eq:calRring}) into (\ref{eq:breveR}), we obtain
\begin{equation} \label{eq:breveRR}
\breve{\cal R}^{\mu \nu \rho} = \breve{R}^{\mu \nu \rho} + 2\kappa \breve{\theta}^{\mu \alpha} \breve{\theta}^{\nu \beta} \breve{\theta}^{\rho \gamma} {\cal \breve{B}}_{\alpha \beta \gamma} \, ,
\end{equation}
where
\begin{equation}
\breve{R}^{\mu \nu \rho} = \mathring{ R}^{\alpha \beta \gamma} ({\cal C}^{-1})^\mu_{\ \alpha}  ({\cal C}^{-1})^\nu_{\ \beta}  ({\cal C}^{-1})^\rho_{\ \gamma}
\end{equation}
The $\breve{R}$-flux can further be rewritten by 
\begin{eqnarray} \label{eq:Rbreve}
\breve{R}^{\mu \nu \rho}&=& ({\cal C}^2 \breve{\theta})^{\alpha \sigma} \partial_\sigma ({\cal C}^2 \breve{\theta})^{\beta \gamma} ({\cal C}^{-1})^\mu_{\ \alpha} ({\cal C}^{-1})^\nu_{\ \beta} ({\cal C}^{-1})^\rho_{\ \gamma} + cyclic \\ \notag
&=& ({\cal C} \breve{\theta})^{\mu \sigma} \partial_\sigma \breve{\theta}^{\nu \rho} - ({\cal C} \breve{\theta})^{\mu \sigma}({\cal C} \breve{\theta})^{\beta \gamma} (\partial_\sigma ({\cal C}^{-1})^\nu_{\ \beta} ({\cal C}^{-1})^\rho_{\ \gamma} +({\cal C}^{-1})^\nu_{\ \beta}\partial_\sigma ({\cal C}^{-1})^\rho_{\ \gamma}) + cyclic\\ \notag
&=& \breve{\theta}^{\mu \sigma} \hat{\partial}_\sigma \breve{\theta}^{\nu \rho} +\breve{\theta}^{\mu \alpha} \breve{\theta}^{\rho \beta} \hat{\partial_\alpha}({\cal C}^{-1})^\nu_{\ \tau} {\cal C}^\tau_{\ \beta}-\breve{\theta}^{\mu \beta} \breve{\theta}^{\nu \alpha} \hat{\partial}_\beta ({\cal C}^{-1})^\rho_{\ \tau} {\cal C}^\tau_{\ \alpha}+ cyclic \\ \notag
&=& \breve{\theta}^{\mu \sigma} \hat{\partial}_\sigma \breve{\theta}^{\nu \rho} + \breve{\theta}^{\nu \sigma} \hat{\partial}_\sigma \breve{\theta}^{\rho \mu}+ \breve{\theta}^{\rho \sigma} \hat{\partial}_\sigma \breve{\theta}^{\mu \nu} \\ \notag
&&-\breve{\theta}^{\mu \alpha} \breve{\theta}^{\rho \beta} ({\cal C}^{-1})^\nu_{\ \tau} \hat{\partial_\alpha}{\cal C}^\tau_{\ \beta}+\breve{\theta}^{\mu \beta} \breve{\theta}^{\nu \alpha} ({\cal C}^{-1})^\rho_{\ \tau} \hat{\partial}_\beta {\cal C}^\tau_{\ \alpha} -\breve{\theta}^{\nu \alpha} \breve{\theta}^{\mu \beta} ({\cal C}^{-1})^\rho_{\ \tau} \hat{\partial_\alpha}{\cal C}^\tau_{\ \beta}\\ \notag 
&& +\breve{\theta}^{\nu \beta} \breve{\theta}^{\rho \alpha} ({\cal C}^{-1})^\mu_{\ \tau} \hat{\partial}_\beta {\cal C}^\tau_{\ \alpha}-\breve{\theta}^{\rho \alpha} \breve{\theta}^{\nu \beta} ({\cal C}^{-1})^\mu_{\ \tau} \hat{\partial_\alpha}{\cal C}^\tau_{\ \beta}+\breve{\theta}^{\rho \beta} \breve{\theta}^{\mu \alpha} ({\cal C}^{-1})^\nu_{\ \tau} \hat{\partial}_\beta {\cal C}^\tau_{\ \alpha}\, \\ \notag
&=& \breve{\theta}^{\mu \sigma} \hat{\partial}_\sigma \breve{\theta}^{\nu \rho} + \breve{\theta}^{\nu \sigma} \hat{\partial}_\sigma \breve{\theta}^{\rho \mu}+ \breve{\theta}^{\rho \sigma} \hat{\partial}_\sigma \breve{\theta}^{\mu \nu}- (\breve{\theta}^{\mu \alpha} \breve{\theta}^{\rho \beta} \breve{f}_{\!\alpha \beta}^{\ \nu} + \breve{\theta}^{\nu \alpha} \breve{\theta}^{\mu \beta} \breve{f}_{\!\alpha \beta}^{\ \rho}+\breve{\theta}^{\rho \alpha} \breve{\theta}^{\nu \beta} \breve{f}_{\!\alpha \beta}^{\ \mu}) \, .
\end{eqnarray}
First, we expressed the flux as a function of $\breve{\theta}$ according to equation (\ref{eq:brthdef}). Then, we applied the chain rule and the derivative $\hat{\partial}$ (\ref{eq:hatpartial}). Finally, we applied the chain rule again to hyperbolic functions and used equation (\ref{eq:brevef}) to obtain the final expression.

Lastly, the remaining algebra between currents is obtained from (\ref{eq:iotakring}) and (\ref{eq:aux-C}) in a following way 
\begin{eqnarray} \label{eq:iota-kappa-0} \notag
\{ \breve{\iota}_\mu(\sigma) , \breve{k}^\nu (\bar{\sigma})  \} &=& {\cal C}^\sigma_{\ \mu} ({\cal C}^{-1})^\nu_{\ \rho} \{ \mathring{\iota}_\sigma ,  \mathring{k}^\rho (\bar{\sigma})   \} +  {\cal C}^\sigma_{\ \mu}\{ \mathring{\iota}_\sigma , ({\cal C}^{-1})^\nu_{\ \rho} \} \mathring{k}^\rho   +  \mathring{\iota}_\sigma \{ {\cal C}^\sigma_{\ \mu},  \mathring{k}^\rho   \} ({\cal C}^{-1})^\nu_{\ \rho} \\ 
&=& \kappa {\cal C}^\sigma_{\ \mu} (\sigma) ({\cal C}^{-1})^\nu_{\ \sigma} (\bar{\sigma})\delta^{\prime}(\sigma-\bar{\sigma})\\ \notag 
&&  + \Big( {\cal \mathring{F}}^{\ \sigma}_{\!\!\rho \tau}\ {\cal C}^\rho_{\ \mu} ({\cal C}^{-1})^\nu_{\ \sigma}-\hat{\partial}_\mu({\cal C}^{-1})^\nu_{\ \tau}  \Big) \mathring{k}^\tau \delta(\sigma-\bar{\sigma}) \\ \notag
&& +\Big(-\kappa {\cal \mathring{Q}}_{\rho}^{\ \sigma \tau} {\cal C}^\rho_{\ \mu} ({\cal C}^{-1})^\nu_{\ \sigma} + \kappa ({\cal C}^{-1})^\nu_{\ \sigma} \mathring{\theta}^{\sigma \rho} \partial_\rho {\cal C}^\tau_{\ \mu} \Big) \mathring{\iota}_\tau \delta(\sigma-\bar{\sigma}) \, .
\end{eqnarray}
The anomalous term becomes
\begin{eqnarray}\label{eq:CC-anom}
\kappa {\cal C}^\sigma_{\ \mu} (\sigma) ({\cal C}^{-1})^\nu_{\ \sigma} (\bar{\sigma})\delta^{\prime}(\sigma-\bar{\sigma})  &=&  \kappa \delta^\nu_\mu \delta^{\prime}(\sigma-\bar{\sigma}) +\kappa {\cal C}^\sigma_{\ \mu} \partial_\rho   ({\cal C}^{-1})^\nu_{\ \sigma} x^{\prime \rho} \delta(\sigma-\bar{\sigma})  \\  \notag
&=& \kappa \delta^\nu_\mu \delta^{\prime}(\sigma-\bar{\sigma})  + {\cal C}^\rho_{\ \mu} \partial_\sigma ({\cal C}^{-1})^\nu_{\ \rho} \mathring{k}^\sigma \delta(\sigma-\bar{\sigma}) \\ \notag
&&  - \kappa {\cal C}^\rho_{\ \mu} \partial_\sigma ({\cal C}^{-1})^\nu_{\ \rho}\mathring{\theta}^{\sigma \tau} \mathring{\iota}_\tau \delta(\sigma-\bar{\sigma}) \, ,
\end{eqnarray}
where we used the well-known delta function identity 
\begin{equation} \label{eq:delta}
f(\bar{\sigma}) \delta^{\prime}(\sigma-\bar{\sigma}) = f^{\prime}(\sigma)\delta(\sigma-\bar{\sigma}) + f(\sigma) \delta^{\prime}(\sigma-\bar{\sigma}) \, .
\end{equation}
Substituting (\ref{eq:CC-anom}) into (\ref{eq:iota-kappa-0}), we obtain 
\begin{eqnarray} \label{eq:iota-kappa-1}
\{ \breve{\iota}_\mu(\sigma) , \breve{k}^\nu (\bar{\sigma})  \}  &=& \kappa \delta^\nu_\mu \delta^{\prime}(\sigma-\bar{\sigma})\\ \notag
&&+\Big({\cal C}^\rho_{\ \mu}(\partial_\sigma ({\cal C}^{-1})^\nu_{\ \rho} - \partial_\rho ({\cal C}^{-1})^\nu_{\ \sigma} ) + {\cal C}^\rho_{\ \mu} ({\cal C}^{-1})^\nu_{\ \tau} \mathring{{\cal F}}_{\!\!\rho \sigma}^{\ \tau} \Big) \mathring{k}^\sigma \delta(\sigma-\bar{\sigma}) \\  \notag
&&+\kappa \Big(({\cal C}^{-1})^\nu_{\ \sigma} \partial_\rho {\cal C}^\tau_{\ \mu}\mathring{\theta}^{\sigma \rho}- {\cal C}^\rho_{\ \mu} \partial_\sigma ({\cal C}^{-1})^\nu_{\ \rho}\mathring{\theta}^{\sigma \tau}\\ \notag 
&& -{\cal C}^\rho_{\ \mu}({\cal C}^{-1})^\nu_{\ \sigma}{\cal \mathring{Q}}_\rho^{\ \sigma \tau} \Big) \mathring{\iota}_\tau \delta(\sigma-\bar{\sigma}) \, .
\end{eqnarray}
Substituting relations between currents (\ref{eq:ik-ik}) in the previous expression, we obtain
\begin{eqnarray} \label{eq:iota-kappa}
\{ \breve{\iota}_\mu(\sigma) , \breve{k}^\nu (\bar{\sigma})  \} = \kappa \delta^\nu_\mu \delta^{\prime}(\sigma-\bar{\sigma})+ \breve{\cal F}^{\ \nu}_{\!\!\mu \rho}\ \breve{k}^\rho \delta(\sigma-\bar{\sigma}) - \kappa \breve{\cal Q}_{\mu}^{\ \nu \rho} \breve{\iota}_\rho  \delta(\sigma-\bar{\sigma}) \, ,
\end{eqnarray}

All Poisson bracket relations between currents (\ref{eq:iota-iota}), (\ref{eq:kappa-kappa}), (\ref{eq:iota-kappa}) can be summarized by
\begin{equation}
\{\breve{X}^M , \breve{X}^N \} = -\breve{F}^{MN}_{\ \ \ \ P}\ \breve{X}^P \delta(\sigma-\bar{\sigma}) + \kappa \eta^{MN} \delta^{\prime}(\sigma-\bar{\sigma}) \, ,
\end{equation}
where
\begin{eqnarray}\label{eq:breveFdef}
 \breve{F}^{M N \rho} =
\begin{pmatrix}
  \kappa^2  \breve{{\cal R}}^{\mu \nu \rho}  & - \kappa \breve{{\cal Q}}_\nu^{\ \mu \rho} \\
\kappa \breve{{\cal Q}}_\mu^{\ \nu \rho}   &  \breve{{\cal F}}^{\ \rho}_{\!\!\mu \nu}  \\
\end{pmatrix}  \,  , \qquad
 \breve{F}^{M N}{}_\rho  =
\begin{pmatrix}
  \kappa \breve{{\cal Q}}_\rho^{\ \mu \nu}   &  \breve{{\cal F}}^{\ \mu}_{\!\!\nu \rho }  \\
 -\breve{{\cal F}}^{\ \nu}_{\!\!\mu \rho}   &   2 {\cal \breve{B}}_{\mu \nu \rho} \\
\end{pmatrix} \, .
 \end{eqnarray}
We are now going to consider structures that naturally appear in the investigation of the Courant bracket simultaneously twisted by $B$ and $\theta$ and its fluxes. 

\section{Twisted Lie bracket}

The matrix ${\cal C}$ appears in all relations as a consequence of simultaneous twisting. Therefore, we start with seeking the Lie algebroid that consists of the tangent bundle, and the anchor is defined with the matrix ${\cal C}$, defined by a simple multiplication ${\cal C} (\xi) = {\cal C} \xi$. The bracket of this Lie algebroid should be related to the Lie bracket by
\begin{eqnarray}
\Big( {\cal C} [\xi_1, \xi_2]_{\hat{L}} \Big)^\mu &=&  \Big( [{\cal C} \xi_1, {\cal C} \xi_2]_{L} \Big)^\mu =  {\cal C}^\nu_{\ \rho} \xi_1^\rho \partial_\nu ({\cal C}^\mu_{\ \sigma} \xi_2^\sigma)- {\cal C}^\nu_{\ \rho} \xi_2^\rho \partial_\nu ({\cal C}^\mu_{\ \sigma} \xi_1^\sigma)\\ \notag
&=& {\cal C}^\nu_{\ \rho} {\cal C}^\mu_{\ \sigma} \Big(\xi_1^\rho \partial_\nu \xi_2^\sigma - \xi_2^\rho \partial_\nu \xi_1^\sigma\Big) + \xi_1^\rho \xi_2^\sigma \Big({\cal C}^\nu_{\ \rho} \partial_\nu {\cal C}^\mu_{\ \sigma}-{\cal C}^\nu_{\ \sigma} \partial_\nu {\cal C}^\mu_{\ \rho} \Big) \\ \notag
&=&{\cal C}^\mu_{\ \sigma}\Big( \xi_1^\rho \hat{\partial}_\rho \xi_2^\sigma -\xi_2^\rho \hat{\partial}_\rho \xi_1^\sigma\Big) +\xi_1^\rho \xi_2^\sigma \Big(\hat{\partial}_\rho {\cal C}^\mu_{\ \sigma}-\hat{\partial}_\sigma{\cal C}^\mu_{\ \rho} \Big)\, , 
\end{eqnarray}
where we used (\ref{eq:hatpartial}) and relabeled some indices. Multiplying the previous relation with ${\cal C}^{-1}$ and taking into the account (\ref{eq:brevef}), we obtain
\begin{equation} \label{eq:hatLie}
\Big( [\xi_1, \xi_2]_{\hat{L}} \Big)^\mu = \xi_1^\nu \hat{\partial}_\nu \xi_2^\mu - \xi_2^\nu \hat{\partial}_\nu \xi_1^\mu + \breve{f}^{\ \mu}_{\!\nu \rho}\ \xi_1^\nu \xi_2^\rho \, .
\end{equation}
Analogous to our notation for twisted Courant brackets, we will denote this bracket as the twisted Lie bracket, since 
\begin{equation} \label{eq:twisted-lie-def}
[\xi_1 ,\xi_2]_{\hat{L}} = {\cal C}^{-1}[{\cal C}\xi_1, {\cal C} \xi_2]_L \, .
\end{equation}
We can now see the interpretation of $\breve{f}$-flux - it is a direct consequence of non-commutativity of partial derivatives under the twisted Lie bracket. Taking the twisted Lie bracket between two partial derivatives, one obtains 
\begin{equation} \label{eq:f-twisted-def}
[\partial_\mu, \partial_\nu]_{\hat{L}}= \breve{f}_{\!\mu \nu}^{\ \rho}\ \partial_\rho \, ,
\end{equation}
and therefore, the $\breve{f}$-flux represent the structure functions of the twisted Lie bracket between two holonomic partial derivatives. 

What does the twisted Lie bracket represent? Again, consider a case of constant background fields $B$ and $\theta$, resulting in constant matrix ${\cal C}$. In this instance, recall that we can define proper coordinate system $\hat{x}^\mu$, such that $\hat{\partial}_\mu = \frac{\partial}{\partial \hat{x}^{\mu}}$, and the matrix ${\cal C}$ is the Jacobiator. Then, this is just a push-forward $\Phi_{\star, p}: T_p{\cal M} \to T_{\Phi(p)}{\cal M}$ of a transformation $\Phi: {\cal M} \to {\cal M}$ where $\partial_\mu \Phi^\nu =({\cal C}^T)_\mu^{\ \nu}$, which can be integrated. As such, relation (\ref{eq:twisted-lie-def}) can be rewritten as 
\begin{equation}
\left. \Phi_{\star} [\xi_1, \xi_2]_L \right| _{p}  = \left. [\Phi_{\star} \xi_1, \Phi_{\star} \xi_2]_L \right| _{\Phi(p)}\, ,
\end{equation}
which is just the well known property that the push-forward of Lie bracket is the Lie bracket of push-forward. In a more general case, the twisted Lie bracket cannot be seen as the push-forward of the (ordinary) Lie bracket, and we have a completely new bracket.

One can be tempted to think of the twisted Lie bracket as the Lie bracket rewritten in a non-holonomic basis spanned by $\hat{\partial}_\mu$. This is however not the case, as can be easily verified
\begin{eqnarray}\label{eq:hatLie1}
[\xi_1, \xi_2]_L &=& \hat{\xi}_1^\mu \hat{\partial}_\mu (\hat{\xi}_2^\nu \hat{\partial}_\nu)-\hat{\xi}_1^\mu \hat{\partial}_\mu (\hat{\xi}_2^\nu \hat{\partial}_\nu) \\ \notag
&=& (\hat{\xi}_1^\nu \hat{\partial}_\nu \hat{\xi}_2^\mu - \hat{\xi}_2^\nu \hat{\partial}_\nu \hat{\xi}_1^\mu + \breve{f}^{\ \mu}_{\!\nu \rho}\ \hat{\xi}_1^\nu \hat{\xi}_2^\rho) \hat{\partial}_\mu \, .
\end{eqnarray}
We obtain the same form as in (\ref{eq:hatLie}), but the components in the expression (\ref{eq:hatLie1}) are expressed in basis $\hat{\partial}_\mu$, while in (\ref{eq:hatLie}) the basis is $\partial_\mu$. 

In order for ${\cal C}$ to be a proper anchor of a Lie algebroid, we have to demonstrate that the Leibniz rule and Jacobi identity are satisfied. Firstly, using definition of the twisted Lie bracket (\ref{eq:twisted-lie-def}) together with Leibniz rule for Lie bracket, we obtain
\begin{eqnarray} \label{eq:hat-lie-alg-1}
[\xi_1, f \xi_2]_{\hat{L}} &=& {\cal C}^{-1}[{\cal C}\xi_1, f {\cal C} \xi_2]_L = {\cal C}^{-1}\Big(({\cal L}_{{\cal C} \xi_1} f )\ \xi_2 + f[{\cal C} \xi_1, {\cal C} \xi_2]_L \Big) \\ \notag
&=& ({\cal L}_{{\cal C} \xi_1} f )\ \xi_2 + f [\xi_1, \xi_2]_{\hat{L}} \, ,
\end{eqnarray}
and therefore we proved (\ref{eq:lialg-def2}). The Jacobi identity (\ref{eq:lialg-def3}) is also satisfied, since
\begin{equation}
[\xi_1, [\xi_2, \xi_3]_{\hat{L}}]_{\hat{L}} + cyclic = {\cal C}^{-1}[{\cal C}\xi_1, [{\cal C}\xi_2, {\cal C}\xi_3]_L]_L + cyclic = {\cal C}^{-1} [\xi_1, [\xi_2, \xi_3]_L]_L + cyclic = 0 \, ,
\end{equation}
and therefore the tangent bundle equipped with the twisted Lie bracket indeed defines a Lie algebroid with ${\cal C}$ as its anchor.

From the Leibniz rule for twisted Lie bracket, we can derive the action of corresponding Lie derivative on functions
\begin{equation}  \label{eq:hat-lie-f}
{\cal \hat{L}}_{\xi}\ f = {\cal L}_{{\cal C}\xi}\ f = \xi^\mu \hat{\partial}_\mu f \, .
\end{equation}
Its action on vectors is simply given by the twisted Lie bracket. 
To write the action of Lie derivative ${\cal L}_{\hat{\xi}}$ on 1-forms, we firstly apply the Leibniz rule (\ref{eq:hat-lie-alg-1}) on 1-form-vector contraction
\begin{eqnarray}
{\cal \hat{L}}_{\xi_1} (\xi_2^\mu \lambda_{2\mu}) &=& ({\cal \hat{L}}_{\xi_1} \xi_2)^\mu \lambda_{2\mu}+\xi_2^\mu ({\cal \hat{L}}_{\xi_1} \lambda_2)_{\mu} \\ \notag
&=&  (\xi_1^\nu \hat{\partial}_\nu \xi_2^\mu-\xi_2^\nu \hat{\partial}_\nu \xi_1^\mu )\lambda_{2\mu} + \breve{f}^{\ \mu}_{\!\nu \rho}\ \xi_1^\nu \xi_2^\rho \lambda_{2\mu}+ \xi_2^\mu  ({\cal \hat{L}}_{\xi_1} \lambda_2)_{\mu} \, ,
\end{eqnarray}
and then (\ref{eq:hat-lie-f}) on that contraction, since it is effectively a scalar
\begin{eqnarray}
{\cal \hat{L}}_{\xi_1} (\xi_2^\mu \lambda_{2\mu})  = \xi_1^\nu \hat{\partial}_\nu (\xi_2^\mu \lambda_{2\mu}) = \xi_1^\nu \hat{\partial}_\nu \xi_2^\mu \lambda_{2\mu} + \xi_1^\nu \xi_2^\mu \hat{\partial}_\nu \lambda_{2\mu}\, .
\end{eqnarray}
When we equate right-hand sides of previous two relations, we obtain
\begin{equation} \label{eq:hat-lie-form}
 ({\cal \hat{L}}_{\xi_1} \lambda_2)_\mu = \hat{\partial}_\mu \xi_1^\nu \lambda_{2\nu} +\xi_1^\nu \hat{\partial}_\nu \lambda_{2\mu} +\breve{f}^{\ \rho}_{\!\mu \nu}\ \xi_1^\nu \lambda_{2\rho} = \xi_1^\nu(\hat{\partial}_\nu \lambda_{2\mu} - \hat{\partial}_\mu \lambda_{2\nu}) + \hat{\partial}_\mu (\xi_1^\nu \lambda_{2\nu}) +\breve{f}^{\ \rho}_{\!\mu \nu}\ \xi_1^\nu \lambda_{2\rho} \, .
\end{equation}
The exterior algebra is easily derived from the relation (\ref{eq:ext-der-rho}). Let us explicitly obtain the action of exterior derivative on functions and 1-forms. Functions correspond to the case $p=0$, so we have
\begin{equation} \label{eq:hat-der-f}
\hat{d} f (\xi) = {\cal C} \xi (f) = \xi^\mu \hat{\partial}_\mu f \, , \quad (\hat{d}f)_{\mu} = \hat{\partial}_\mu f \, .
\end{equation}
From (\ref{eq:hat-lie-f}), we see that the usual relation for the action of Lie derivatives on function ${\cal \hat{L}}_{\xi} f = i_{\xi} \hat{d} f$ still holds in the twisted case.
On the other hand, 1-forms correspond to the case of $p=1$ in (\ref{eq:ext-der-rho}), from which we obtain
\begin{eqnarray} \label{eq:hat-der-lambda}
\hat{d} \lambda (\xi_1, \xi_2) &=& {\cal C}\xi_1 (\lambda (\xi_2))-{\cal C}\xi_2 (\lambda (\xi_1))- \lambda([\xi_1, \xi_2]_{\hat{L}}) \\ \notag
&=& \xi_1^\mu \xi_2^\nu \Big(\hat{\partial}_\mu \lambda_\nu - \hat{\partial}_\nu \lambda_\mu - \hat{f}^{\ \rho}_{\!\mu \nu} \lambda_\rho \Big) \, , \\ \notag
(\hat{d} \lambda)_{\mu \nu} &=& \hat{\partial}_\mu \lambda_\nu - \hat{\partial}_\nu \lambda_\mu - \hat{f}^{\ \rho}_{\!\mu \nu}\ \lambda_\rho \, .
\end{eqnarray}
The Cartan formula ${\cal \hat{L}}_{\xi} \lambda = i_{\xi} \hat{d}\lambda + \hat{d}i_{\xi} \lambda $ can be easily demonstrated using (\ref{eq:hat-lie-form}) and (\ref{eq:hat-der-lambda}), and holds true for any $p$-form $\lambda$.

\subsection{Generalized H-flux}

Let us now reconsider the generalized $\breve{H}$-flux (\ref{eq:breveB}) in terms of new Lie algebroid that we obtained. It has a structure of a 3-form, that when contracted with three vectors can be written by
\begin{equation}
\breve{{\cal B}}_{\mu \nu \rho} \xi_1^\mu \xi_2^\nu \xi_3^\rho =  \mathring{B}_{\alpha \beta \gamma}{\cal C}^\alpha_{\ \mu}\xi_1^\mu {\cal C}^\beta_{\ \nu}\xi_2^\nu {\cal C}^\gamma_{\ \rho}\xi_3^\rho  = d \mathring{B}({\cal C}\xi_1, {\cal C}\xi_2, {\cal C}\xi_3)  \, .
\end{equation}
where $\mathring{B}_{\mu \nu \rho}$ is defined in (\ref{eq:fluxring}). The right-hand side of the previous relation is expressed in a non coordinate notation, which using the definition for exterior derivative can be further transformed by
\begin{eqnarray}
(d \mathring{B}) ({\cal C}\xi_1, {\cal C}\xi_2, {\cal C}\xi_3)  &=&  {\cal C}\xi_1 \Big(\mathring{B}({\cal C}\xi_2,{\cal C} \xi_3) \Big)- {\cal C}\xi_2\Big( \mathring{B}({\cal C}\xi_1,{\cal C} \xi_3) \Big)+{\cal C}\xi_3\Big( \mathring{B}({\cal C}\xi_1,{\cal C} \xi_2)\Big) \\ \notag
&&- \mathring{B}\Big([{\cal C}\xi_1, {\cal C}\xi_2]_L, {\cal C}\xi_3 \Big)+\mathring{B}\Big([{\cal C}\xi_1, {\cal C}\xi_3]_L, {\cal C}\xi_2 \Big)-\mathring{B}\Big([{\cal C}\xi_2, {\cal C}\xi_3]_L, {\cal C}\xi_1 \Big) \\ \notag
&=&{\cal C}\xi_1 \Big(\hat{B}(\xi_2, \xi_3) \Big)- {\cal C}\xi_2\Big( \hat{B}(\xi_1,\xi_3) \Big)+{\cal C}\xi_3\Big( \hat{B}(\xi_1,\xi_2)\Big) \\ \notag
&&- \hat{B}\Big([\xi_1, \xi_2]_{\hat{L}}, \xi_3 \Big)+\hat{B}\Big([\xi_1,\xi_3]_{\hat{L}}, \xi_2 \Big)-\hat{B}\Big([\xi_2, \xi_3]_{\hat{L}}, \xi_1 \Big) \\ \notag
&=& \hat{d} \hat{B} (\xi_1, \xi_2, \xi_3) \, ,
\end{eqnarray}
where $\hat{B}$ is a new field that we defined by
\begin{equation} \label{eq:hat-B-def}
\hat{B}_{\mu \nu} = \mathring{B}_{\alpha \beta} {\cal C}^\alpha_{\ \mu} {\cal C}^\beta_{\ \nu} = (B {\cal S C})_{\mu \nu} \, ,
\end{equation}
and in the last step recognized the expression for twisted exterior derivative acting on a 2-form. 

\section{Twisted Koszul bracket}
\cleq 

We define the twisted Koszul bracket by
\begin{equation} \label{eq:twist-koszul}
[\lambda_1, \lambda_2]_{\breve{\theta}} = {\cal \hat{L}}_{\breve{\theta}(\lambda_1)} \lambda_2- {\cal \hat{L}}_{\breve{\theta}(\lambda_2)} \lambda_1 - \hat{d}(\breve{\theta}(\lambda_1, \lambda_2)) \, ,
\end{equation}
where $\breve{\theta}(\lambda_1)^\mu = \lambda_{1\nu}\breve{\theta}^{\nu \mu}$. This is an analogous definition to the one for the (non-twisted) Koszul bracket (\ref{eq:koszul-def}). In some local basis, its components are given by
\begin{eqnarray} \label{eq:twist-koszul-coord}\notag
\Big([\lambda_1, \lambda_2]_{\breve{\theta}} \Big)_\mu &=& \breve{\theta}^{\nu \rho}(\lambda_{1\nu} \hat{\partial}_\rho \lambda_{2\mu}-\lambda_{2\nu} \hat{\partial}_\rho \lambda_{1\mu}) + (\hat{\partial}_\mu \breve{\theta}^{\nu \rho} +\breve{f}^{\ \nu}_{\!\mu \sigma}\ \breve{\theta}^{\sigma \rho}-\breve{f}^{\ \rho}_{\!\mu \sigma}\ \breve{\theta}^{\sigma \nu} )\lambda_{1\nu} \lambda_{2\rho} \\ 
&=& \breve{\theta}^{\nu \rho}(\lambda_{1\nu} \hat{\partial}_\rho \lambda_{2\mu}-\lambda_{2\nu} \hat{\partial}_\rho \lambda_{1\mu}) +\breve{Q}_\mu^{\ \nu \rho}\ \lambda_{1\nu} \lambda_{2\rho}  \, .
\end{eqnarray}
%\hat{\partial}_\rho \breve{\theta}^{\mu \nu} + \breve{f}^{\ \mu}_{\!\rho \sigma}\  \breve{\theta}^{\sigma \nu} - \breve{f}^{\ \nu}_{\!\rho \sigma}\ \breve{\theta}^{\sigma \mu}
Taking the twisted Koszul bracket between one-forms $dx^\mu$, one obtains
\begin{equation} \label{eq:Q-twisted-def}
[dx^\mu, dx^\nu]_{\breve{\theta}} = \breve{Q}_\rho^{\ \mu \nu}\ dx^\rho \, , 
\end{equation}
where we used (\ref{eq:twist-koszul-coord}) and (\ref{eq:Qbreve}). In an analogous way that the $\breve{f}$-flux was defined in terms of twisted Lie bracket (\ref{eq:f-twisted-def}), the $\breve{Q}$-flux is defined as the structure functions of the twisted Koszul bracket of the covectors $dx^\mu$.

The twisted Koszul bracket can be related to the twisted Lie bracket. In order to do that, we firstly obtain 
\begin{eqnarray}
\breve{\theta} \Big([\lambda_1, \lambda_2]_{\breve{\theta}} \Big) &=& \breve{\theta}^{\nu \mu} \Big(\breve{\theta}^{\sigma \rho}(\lambda_{1\sigma } \hat{\partial}_\rho \lambda_{2\nu}-\lambda_{2\sigma} \hat{\partial}_\rho \lambda_{1\nu}) + \hat{\partial}_\nu \breve{\theta}^{\sigma \rho} \lambda_{1\sigma} \lambda_{2\rho}  \\ \notag
&& +(\breve{f}^{\ \rho}_{\!\nu \tau}\ \breve{\theta}^{\tau \sigma}-\breve{f}^{\ \sigma}_{\!\nu \tau}\ \breve{\theta}^{\tau \rho} )\lambda_{1\rho} \lambda_{2\sigma}\Big) \, .
\end{eqnarray}
Next, we obtain
\begin{eqnarray}
[\breve{\theta}(\lambda_1), \breve{\theta}(\lambda_2)]_{\hat{L}} &=& \breve{\theta}^{\nu \rho}\lambda_{1\rho} \hat{\partial}_\nu (\breve{\theta}^{\mu \sigma}\lambda_{2\sigma} )-\breve{\theta}^{\nu \rho}\lambda_{2\rho} \hat{\partial}_\nu (\breve{\theta}^{\mu \sigma}\lambda_{1\sigma} )+ \breve{f}^{\ \mu}_{\!\nu \rho}\ \breve{\theta}^{\nu \sigma} \breve{\theta}^{\rho \tau} \lambda_{1\sigma} \lambda_{2\tau} \\ \notag
&=&\breve{\theta}^{\nu \rho} \breve{\theta}^{\mu \sigma}\Big(\lambda_{1\rho}\hat{\partial}_\nu \lambda_{2\sigma}-\lambda_{2\rho}\hat{\partial}_\nu \lambda_{1\sigma} \Big)+\breve{f}^{\ \mu}_{\!\nu \rho}\ \breve{\theta}^{\nu \sigma}\breve{\theta}^{\rho \tau} \lambda_{1\sigma} \lambda_{2\tau} \\ \notag
&&+(\breve{\theta}^{\rho \nu }\hat{\partial}_\nu \breve{\theta}^{\sigma \mu}+\breve{\theta}^{\sigma \nu}\hat{\partial}_\nu \breve{\theta}^{\mu \rho}) \lambda_{1\rho}\lambda_{2\sigma}\, .
\end{eqnarray}
After relabeling of some dummy indices, we obtain the relation 
\begin{equation}
\Big[ \breve{\theta}  \Big( [\lambda_1, \lambda_2]_{\breve{\theta}} )\Big) \Big]^\mu = \Big( [\breve{\theta}(\lambda_1), \breve{\theta}(\lambda_2)]_{\hat{L}}\Big)^\mu + \breve{R}^{\mu \nu \rho}\lambda_{1\nu} \lambda_{2\rho} \, ,
\end{equation}
where $\breve{R}$ is defined in (\ref{eq:Rbreve}). We can use the definition of the twisted Lie bracket (\ref{eq:twisted-lie-def}), to rewrite the previous relation as
\begin{equation}
\Big[ {\cal C}\breve{\theta} \Big([\lambda_1, \lambda_2]_{\breve{\theta}} \Big) \Big]^\mu   = \Big( [{\cal C}\breve{\theta}(\lambda_1), {\cal C} \breve{\theta}(\lambda_2)]_{L} \Big)^\mu + {\cal C}^{\mu}_{\ \sigma}\breve{R}^{\sigma \nu \rho}\lambda_{1\nu} \lambda_{2\rho} \, .
\end{equation}
The twisted Koszul bracket defines a quasi-Lie algebroid with anchor $\hat{\rho}_{\breve{\theta}} = {\cal C}\breve{\theta}$, where the $\breve{R}$-flux is a deformation from the Lie algebroid structure. 

We can still define the exterior derivative associated with the quasi-Lie algebroid defined with the twisted Koszul bracket. On functions, its action is obtained from (\ref{eq:ext-der-rho})
\begin{equation}
\hat{d}_{\breve{\theta}} f (\lambda) = \breve{\theta}^{\mu \nu} \hat{\partial}_\nu f \lambda_\mu \, , \qquad (\hat{d}_{\breve{\theta}} f)^\mu = \breve{\theta}^{\mu \nu} \hat{\partial}_\nu f \, .
\end{equation}
Similarly, on vectors it becomes 
\begin{eqnarray}
\hat{d}_{\breve{\theta}} \xi (\lambda_1, \lambda_2) &=& (\lambda_{1\rho} \breve{\theta}^{\rho \nu})\hat{\partial}_\nu (\xi^\mu \lambda_{2\mu})-(\lambda_{2\rho} \breve{\theta}^{\rho \nu})\hat{\partial}_\nu (\xi^\mu \lambda_{1\mu})-\xi^\mu \Big( [\lambda_1, \lambda_2]_{\breve{\theta}}\Big)_\mu \\ \notag
&=& \Big( \breve{\theta}^{\mu \rho} \hat{\partial}_\rho \xi^\nu- \breve{\theta}^{\nu \rho} \hat{\partial}_\rho \xi^\mu - \xi^\rho (\hat{\partial}_\rho \breve{\theta}^{\mu \nu}+\breve{f}^{\ \mu}_{\!\rho \sigma}\ \breve{\theta}^{\sigma \nu}-\breve{f}^{\ \nu}_{\!\rho \sigma}\ \breve{\theta}^{\sigma \mu})\Big) \lambda_{1\mu} \lambda_{2\nu} \, .
\end{eqnarray}
The exterior derivative $\hat{d}_{\breve{\theta}}$ satisfies Leibniz rule, but is not idempotent, unless the $\breve{R}$-flux is zero. 

\section{Twisted Schouten-Nijenhuis bracket}
\cleq 

Lastly, let us define the twisted Schouten-Nijenhuis bracket as the generalization of the twisted Lie bracket on the space of multi-vectors
\begin{eqnarray}
[f, g]_{\hat{S}} = 0 \, , \quad [\xi, f]_{\hat{S}} = {\cal \hat{L}}_{\xi}(f) \, , \quad [\xi_1, \xi_2]_{\hat{S}} = [\xi_1, \xi_2]_{\hat{L}} \, ,
\end{eqnarray}
with other brackets obtained via graded relations (\ref{eq:SNB-grad1}) and (\ref{eq:SNB-grad2}). The important feature is that we can express the $\breve{R}$-flux (\ref{eq:Rbreve}) in terms of the twisted Schouten-Nijenhuis bracket via
\begin{equation} \label{eq:breveR-thth}
\breve{R}=\frac{1}{2}[\breve{\theta}, \breve{\theta}]_{\hat{S}}  \, ,
\end{equation}
analogous to their non-twisted counterparts. 

The expression (\ref{eq:breveR-thth}) can be proven using the principle of mathematical induction. We begin by showing the relation (\ref{eq:breveR-thth}) holds for a simple bi-vector $\breve{\theta}$. Next, assuming that the relation (\ref{eq:breveR-thth}) holds for a composite bi-vector consisting of $n$ simple bi-vectors, we proceed to demonstrate that it also holds for the composite bi-vector consisting of $n+1$ simple bi-vectors. While this problem is straightforward to resolve, it involves several technical steps in computations. Detailed derivations for this process are included in the paper's appendix.

\subsection{Relation to the twisted Koszul bracket}

In the previous chapters, we saw that the Schouten-Nijenhuis bracket can be written as $d_{\theta} \theta = [\theta, \theta]_S$. This motivates us to consider the action of exterior derivative $\hat{d}_{\breve{\theta}}$ on the bi-vector $\breve{\theta}$. From definition (\ref{eq:ext-der-rho}), we have
\begin{eqnarray}\label{eq:rho-der-theta-def} \notag
\hat{d}_{\breve{\theta}} \breve{\theta}(\lambda_1, \lambda_2, \lambda_3) &=& {\cal C}\breve{\theta} (\lambda_1)\Big([\lambda_2, \lambda_3]_{\breve{\theta}}\Big)- {\cal C}\breve{\theta} (\lambda_2) \Big([\lambda_1, \lambda_3]_{\breve{\theta}}\Big)+{\cal C}\breve{\theta} (\lambda_3) \Big([\lambda_1, \lambda_2]_{\breve{\theta}}\Big)\\ 
&&-\breve{\theta}\Big([\lambda_1, \lambda_2]_{\breve{\theta}}, \lambda_3 \Big)+\breve{\theta}\Big([\lambda_1, \lambda_3]_{\breve{\theta}}, \lambda_2 \Big)-\breve{\theta}\Big([\lambda_2, \lambda_3]_{\breve{\theta}}, \lambda_1 \Big) \, .
\end{eqnarray}
There are two types of terms, so let us calculate the components of a representative of each type. Firstly, we have
\begin{eqnarray} \label{eq:rho-der-theta-0}
{\cal C}\breve{\theta} (\lambda_1) \Big([\lambda_2, \lambda_3]_{\breve{\theta}}\Big) &=& \lambda_{1\mu} \breve{\theta}^{\mu \sigma} \hat{\partial}_\sigma \Big(\breve{\theta}^{\nu \rho} \lambda_{2\nu } \lambda_{3\rho} \Big) \\ \notag
&=& \breve{\theta}^{\mu \sigma} \hat{\partial}_\sigma \breve{\theta}^{\nu \rho} \lambda_{1\mu} \lambda_{2\nu} \lambda_{3\rho} + \breve{\theta}^{\mu \sigma} \breve{\theta}^{\nu \rho} \lambda_{1\mu} (\hat{\partial}_\sigma \lambda_{2\nu } \lambda_{3 \rho}+ \lambda_{2\nu } \hat{\partial}_\sigma  \lambda_{3\rho}) \, , 
\end{eqnarray}
and secondly
\begin{eqnarray} \label{eq:rho-der-theta-1}
-\breve{\theta}\Big([\lambda_1, \lambda_2]_{\breve{\theta}}, \lambda_3 \Big) &=& \lambda_{1\mu} \lambda_{2\nu } \lambda_{3 \rho} \Big( \breve{\theta}^{\rho \sigma} \hat{\partial}_\sigma \breve{\theta}^{\mu \nu} + \breve{f}^{\ \mu}_{\!\sigma \tau}\ \breve{\theta}^{\nu \sigma} \breve{\theta}^{\rho \tau}- \breve{f}^{\ \nu}_{\!\sigma \tau}\ \breve{\theta}^{\mu \sigma} \breve{\theta}^{\rho \tau}\Big)\\ \notag
&&- \breve{\theta}^{\mu \rho} \breve{\theta}^{\nu \sigma} (\lambda_{1\nu} \hat{\partial}_\sigma \lambda_{2\mu}-\lambda_{2\nu} \hat{\partial}_\sigma \lambda_{1\mu}) \lambda_{3\rho} \, .
\end{eqnarray}
When (\ref{eq:rho-der-theta-0}) and (\ref{eq:rho-der-theta-1}) are substituted in (\ref{eq:rho-der-theta-def}), we obtain %dodati malo koraka možda
\begin{eqnarray}
\hat{d}_{\breve{\theta}} \breve{\theta}(\lambda_1, \lambda_2, \lambda_3) = 2 \breve{R}^{\mu \nu \rho} \lambda_{1\mu} \lambda_{2\nu} \lambda_{3\rho} \, ,
\end{eqnarray}
which is exactly what we expected. We see that equivalently, the twisted Schouten-Nijenhuis bracket, and consequentially $\breve{R}$-flux, can be defined as the action of the twisted exterior derivative related to the quasi Lie algebroid defined with the twisted Koszul bracket:
\begin{equation} \label{eq:twisted-SN-def}
[\breve{\theta}, \breve{\theta}]_{\hat{S}} = \hat{d}_{\breve{\theta}} \breve{\theta} \, .
\end{equation}

\section{Non-holonomic basis}
\cleq

The fluxes can be written in terms of non-holonomic basis $\hat{\partial}_\mu$. For instance, we have
\begin{equation} \label{eq:NH-f}
[\hat{\partial}_\mu , \hat{\partial}_\nu]_L = \breve{f}_{\mu \nu}^{\ \rho}\ \hat{\partial}_\rho \, ,
\end{equation}
which gives the $\breve{f}$ flux in an exact way that the usual $f$-flux is obtained. The similar relations can be easily obtained for other fluxes.  So what was the value of defining all additional (twisted)-Lie algebroid structures? 

Primarily, we would like to emphasize the way that the fluxes were computed in this paper. We started with the currents that were defined in the basis of holonomic vectors and one-forms, and obtained the $\breve{f}$-flux in it. As such, this flux cannot be understood merely as the consequence of change of basis, but rather as an inherent feature of the simultaneous twisting, characterized with the matrix ${\cal C}$. In addition, the role of non-holonomic basis in physics is usually associated with the appropriate vielbeins that locally simplify the metric tensor. In this instance, the opposite thing is true, the transformation matrix mixes the metric with both the Kalb-Ramond field and non-commutativity parameter. Therefore, we found the twisted Lie algebra description to be more fitting. 

Perhaps more straightforward motivation becomes evident from the full expression of the Courant bracket twisted simultaneously by $B$ and $\theta$. It was obtained in \cite{CBTh} and it is given by 
\begin{equation}
[\Lambda_1, \Lambda_2]_{{\cal C}_{\breve{B}}} = \Lambda \, , \quad \Lambda = \xi \oplus \lambda \, , 
\end{equation}
where
\begin{eqnarray} \label{eq:breveRxi}
\xi &=& {\cal C}^{-1}[ {\cal C}\xi_1,{\cal C}\xi_2]_L - {\cal C}^{-1}[{\cal C}\xi_2,\lambda_1 \kappa {\cal C}\breve{\theta}]_L +{\cal C}^{-1}[{\cal C}\xi_1,\lambda_2 \kappa {\cal C}\breve{\theta}]_L \\ \notag
&&- \Big({\cal L}_{{\cal C}{\xi_1}}(\lambda_2{\cal C}^{-1}) - {\cal L}_{{\cal C}{\xi_2}}(\lambda_1{\cal C}^{-1}) - \frac{1}{2}d(i_{\xi_1}\lambda_2 - i_{\xi_2}\lambda_1)\Big)\kappa {\cal C}\breve{\theta}\\ \notag
&&+ \frac{\kappa^2}{2} {\cal C}^{-1} [{\cal C}^2\breve{\theta},{\cal C}^2\breve{\theta}]_S (\lambda_1 {\cal C}^{-1},\lambda_2 {\cal C}^{-1},.) +2\kappa \breve{\theta}\breve{H}(., \xi_1, \xi_2) \\ \notag
&&-2\wedge^2 \kappa\breve{\theta} \breve{H}(\lambda_1 , ., \xi_2) + 2\wedge^2\kappa\breve{H}(\lambda_2 , ., \xi_1)+2\wedge^3\kappa \breve{\theta}\breve{H}(\lambda_1 ,\lambda_2 ,.) \, ,
\end{eqnarray}
and
\begin{eqnarray} \notag 
\lambda &=& \Big({\cal L}_{{\cal C}{\xi_1}}(\lambda_2{\cal C}^{-1}) - {\cal L}_{{\cal C}{\xi_2}}(\lambda_1{\cal C}^{-1}) - \frac{1}{2}d(i_{\xi_1}\lambda_2 - i_{\xi_2}\lambda_1)\Big){\cal C} - [\lambda_1 {\cal C}^{-1},\lambda_2{\cal C}^{-1}]_{\kappa {\cal C}^2\breve{\theta}}{\cal C} \\ \label{eq:breveRLambda}
&&+2 \breve{H}(\xi_1,\xi_2,.)-2 \kappa \breve{\theta}\breve{H} (\lambda_2, . ,\xi_1)+ 2 \kappa \breve{\theta} \breve{H}(\lambda_1, . ,\xi_2)+2 \wedge^2 \kappa \breve{\theta} \breve{H} (\lambda_1, \lambda_2, .) \, .
\end{eqnarray}

The (\ref{eq:breveRxi}) and (\ref{eq:breveRLambda}) are coordinate-free, and will be equal in holonomic and non-holonomic basis alike. In terms of the brackets that we defined in this paper, these expressions are given by
\begin{eqnarray}
\xi &=& [\xi_1, \xi_2]_{\hat{L}} - \kappa \breve{\theta}\Big({\cal \hat{L}}_{\xi_1} \lambda_2- {\cal \hat{L}}_{\xi_2} \lambda_1 - \frac{1}{2} \hat{d}(i_{\xi_1} \lambda_2-i_{\xi_2} \lambda_1)\Big) \\ \notag
&& +[\xi_1, \kappa \breve{\theta}(\lambda_2)]_{\hat{L}}-[\xi_2, \kappa \breve{\theta}(\lambda_1)]_{\hat{L}} + \frac{\kappa^2}{2} [\breve{\theta}, \breve{\theta}]_{\hat{S}} (\lambda_1, \lambda_2, .)  \\ \notag
&&+ 2\kappa \breve{\theta}\ \hat{d}\hat{B} (. , \xi_1, \xi_2) - 2  \wedge^2 \kappa\breve{\theta}\ \hat{d}\hat{B} (., \lambda_1 , \xi_2)  + 2  \wedge^2 \kappa\breve{\theta}\ \hat{d}\hat{B} (., \lambda_2, \xi_1) + 2\wedge^3 \kappa \breve{\theta}\  \hat{d}\hat{B}(\lambda_1, \lambda_2, .)\, ,
\end{eqnarray}
and 
\begin{eqnarray}
\lambda &=& {\cal \hat{L}}_{\xi_1} \lambda_2- {\cal \hat{L}}_{\xi_2} \lambda_1 + \frac{1}{2} \hat{d}(i_{\xi_1} \lambda_2-i_{\xi_2} \lambda_1)+ \kappa [\lambda_1, \lambda_2]_{\breve{\theta}} \\ \notag
&&+2\hat{d}\hat{B}(\xi_1, \xi_2, .) -2 \kappa \breve{\theta}\ \hat{d}\hat{B} (\lambda_2, . ,\xi_1)+ 2 \kappa \breve{\theta}\ \hat{d}\hat{B} (\lambda_1, . ,\xi_2)+2 \wedge^2 \kappa \breve{\theta} \ \hat{d}\hat{B} (\lambda_1, \lambda_2, .) \, .
\end{eqnarray}
We see that the brackets that we introduced in this paper naturally appear from the bracket we investigated.

\section{Conclusions}
\cleq

In this paper, we investigated the generalized fluxes that appear as structure functions in the Poisson bracket algebra of non-canonical currents, the linear combination of which produces the Courant bracket simultaneously twisted by $B$ and $\theta$. These currents are mutually related by T-duality, which effectively makes the Courant bracket simultaneously twisted by $B$ and $\theta$ into the T-dual invariant extension of the Courant bracket containing all fluxes. In order to obtain these fluxes, we used the relation between this twisted Courant bracket and another twisted Courant bracket that also contains all fluxes \cite{CBTh}. These brackets are related by the hyperbolic function ${\cal C}$ that depends on the product of $B$ and $\theta$ fields. It turned out that in the description of underlining geometries in which fluxes appear this hyperbolic function plays an important role.

Firstly, we considered the Lie algebroid defined on the smooth sections of the tangent bundle, with the requirement that the hyperbolic ${\cal C}$ matrix serves as its anchor. The resulting compatible bracket is the twisted Lie bracket. The algebra of the twisted Lie bracket between two partial derivatives $\partial_\mu$ closes with $\breve{f}$-flux as the structure functions. Generally, the $f$-flux is seen as the so called geometric flux, and appears in the Lie algebra of some non-holonomic basis in which metric is diagonal. In our case, we do not have diagonal metric on the manifold, and moreover, the $f$-flux depends on both $B$ and $\theta$ background fields. Given that a bi-vector is not always defined globally, we cannot talk about the $\breve{f}$-flux as a ''geometric'' flux that is defined globally. Instead, the $\breve{f}$-flux is a direct consequence of simultaneous twisting. 

Secondly, we defined the exterior derivative related to the twisted Lie bracket. We demonstrated that the $\breve{H}$ flux can be seen as the twisted exterior derivative of the new background field $\hat{B}$. This flux is also dependent on both fields $B$ and $\theta$, which differs from analogous fluxes in previous works. We can think of simultaneous twisting of the Courant bracket resulting in the twisting of the $H$-flux as well. 

Thirdly, we defined the twisted Koszul bracket by replacing the standard invariant terms in the expression for (untwisted) Koszul bracket with their twisted counterparts. We showed that when two one-forms $dx^\mu$ are considered, their twisted Koszul bracket produces the $\breve{Q}$-flux. This way, we obtained the analogous relation between the $\breve{f}$-flux and $\breve{Q}$-flux. The former are structure functions in the twisted Lie bracket algebra between holonomic partial derivatives, while the latter are structure functions in the twisted Koszul bracket algebra between their mathematical duals. In addition, we obtained the analogous relation between the usual $Q$-flux and Koszul bracket on one side, and the $\breve{Q}$-flux and the twisted Koszul bracket on the other side.

Lastly, in order to write the $\breve{R}$-flux in coordinate-free form, we introduced the twisted Schouten-Nijenhuis bracket as the graded twisted Lie bracket. The $\breve{R}$-flux can also be obtained from the action of the exterior derivative associated with the twisted Koszul bracket on the bi-vector $\breve{\theta}$.

It is well known that the Koszul bracket is a Lie algebroid bracket only when the bi-vector is of the Poisson type, while in the other case it is the bracket of a quasi-Lie algebroid. In case of the twisted Koszul bracket, the Lie algebroid conditions will be satisfied if the twisted Schouten-Nijenhuis bracket of the bi-vector $\breve{\theta}$ with itself is zero. 

In a trivial case of constant fields $B$ and $\theta$, the matrix ${\cal C}$ is also constant and represent the diffeomorphisms. Obviously, all fluxes are zero in this instance, but it is interesting to see the relations between the newly obtained twisted brackets. The compatibility condition between the anchor and the twisted Lie bracket becomes the push-forward invariance of the Lie bracket. 

The research of Courant algebroids found many applications relevant for physicists. For instance, to each Courant algebroid it is possible to construct a corresponding Courant $\sigma$-model, characterized by topological action of AKSZ type that satisfies classical master equation \cite{CA-sigma}. Physically, this is a theory of a topological brane. This can be used to analyze bosonic string in different flux backgrounds. Moreover, the complete BV-BRST formulation of the Courant $\sigma$-model can be described in terms of $L_{\infty}$ algebras \cite{L-infty}. It would be interesting to explore whether we can gain insights from the membrane described by this Courant algebroid, which encompasses all fluxes and is manifestly invariant under T-duality.

Additionally, backgrounds were considered in which algebra of generalized vielbeins is described by Courant algebroids. For example, this way torus with fluxes obtained by application of T-duality along different directions can be described. We hope that the Courant algebroid with generalized fluxes that we obtained in this paper could also describe some background relevant for string theory, which is a work that remains to be done.

\appendix

\section{Derivation of the relation between the twisted Schouten-Nijenhuis bracket and $\breve{R}$-flux}

In this appendix, we will show that the relation (\ref{eq:breveR-thth}) is indeed correct for all bi-vectors 
\begin{equation}
\breve{\theta} = \sum_{i=1}^{n} \hat{\xi}_{i1} \wedge \hat{\xi}_{i2} \, ,
\end{equation}
using the principle of mathematical induction. For the base of our induction, we will consider the case of a simple bi-vector (for which $n=1$). Then, we will show that if one assumes that the relation (\ref{eq:breveR-thth}) hold for $n$, it will also hold for $n+1$. 

\subsection{Simple bi-vector}

Let us assume that the bi-vector $\breve{\theta}$ is decomposable bi-vector, meaning that it can be expressed as a wedge product between some two vectors $\hat{\xi}_1$ and $\hat{\xi}_2$  
\begin{equation} \label{eq:simple-th}
\breve{\theta} = \hat{\xi}_1 \wedge \hat{\xi}_2 \, , \qquad \breve{\theta}^{\mu \nu} = \hat{\xi}_1^\mu \hat{\xi}_2^\nu - \hat{\xi}_1^\nu \hat{\xi}_2^\mu  \, .
\end{equation}
Using the graded Leibniz identity (\ref{eq:SNB-grad1}), we obtain 
\begin{eqnarray} \label{eq:v-th1}
[\hat{\xi}_1, \breve{\theta}]_{\hat{S}} = [\hat{\xi}_1, \hat{\xi}_1 \wedge \hat{\xi}_2]_{\hat{S}} = [\hat{\xi}_1, \hat{\xi}_1]_{\hat{S}} \wedge \hat{\xi}_2 +\hat{\xi}_1 \wedge [\hat{\xi}_2, \hat{\xi}_1]_{\hat{S}} = \hat{\xi}_1 \wedge [\hat{\xi}_2, \hat{\xi}_1]_{\hat{L}} \, , 
\end{eqnarray}
where we also used the third relation in (\ref{eq:SNB-def}). Similarly, we have
\begin{eqnarray}\label{eq:v-th2}
[\hat{\xi}_2, \breve{\theta}]_{\hat{S}} = -[\hat{\xi}_1, \hat{\xi}_2]_{\hat{L}} \wedge \hat{\xi}_2 \, .
\end{eqnarray}
Now, the application of relations (\ref{eq:SNB-grad1}), (\ref{eq:SNB-grad2}), (\ref{eq:v-th1}) and (\ref{eq:v-th2}), we obtain
\begin{eqnarray} \label{eq:SNB-der}
[\breve{\theta}, \breve{\theta}]_{\hat{S}} = [\breve{\theta}, \hat{\xi}_1]_{\hat{S}} \wedge \hat{\xi}_2 - \hat{\xi}_1 \wedge [\breve{\theta}, \hat{\xi}_2]_{\hat{S}} = - 2 \hat{\xi}_1 \wedge [\hat{\xi}_1, \hat{\xi}_2]_{\hat{L}} \wedge \hat{\xi}_2 \, .
\end{eqnarray}
In local basis spanned by $\partial_\mu$, the components of the twisted Schouten-Nijenhuis bracket are
\begin{eqnarray} \label{eq:SNB-v}
\frac{1}{2}\Big( [\breve{\theta}, \breve{\theta}]_{\hat{S}} \Big)^{\mu \nu \rho} &=& - \sum_{\tiny{odd}} \hat{\xi}_1^\mu ([\hat{\xi}_1, \hat{\xi}_2]_{\hat{L}})^\nu \hat{\xi}_2^\rho  \\ \notag
&=& -\hat{\xi}_1^\mu (\hat{\xi}_1^\sigma \hat{\partial}_\sigma \hat{\xi}_2^\nu - \hat{\xi}_2^\sigma \hat{\partial}_\sigma \hat{\xi}_1^\nu + \breve{f}_{\! \alpha \beta}^{\ \nu}\ \hat{\xi}_1^\alpha \hat{\xi}_2^\beta) \hat{\xi}_2^\rho \\ \notag
&&+\hat{\xi}_1^\rho (\hat{\xi}_1^\sigma \hat{\partial}_\sigma \hat{\xi}_2^\nu - \hat{\xi}_2^\sigma \hat{\partial}_\sigma \hat{\xi}_1^\nu + \breve{f}_{\! \alpha \beta}^{\ \nu}\ \hat{\xi}_1^\alpha \hat{\xi}_2^\beta) \hat{\xi}_2^\mu \\ \notag
&&-\hat{\xi}_1^\nu (\hat{\xi}_1^\sigma \hat{\partial}_\sigma \hat{\xi}_2^\rho - \hat{\xi}_2^\sigma \hat{\partial}_\sigma \hat{\xi}_1^\rho + \breve{f}_{\! \alpha \beta}^{\ \rho}\ \hat{\xi}_1^\alpha \hat{\xi}_2^\beta) \hat{\xi}_2^\mu \\ \notag
&&+\hat{\xi}_1^\mu (\hat{\xi}_1^\sigma \hat{\partial}_\sigma \hat{\xi}_2^\rho - \hat{\xi}_2^\sigma \hat{\partial}_\sigma \hat{\xi}_1^\rho + \breve{f}_{\! \alpha \beta}^{\ \rho}\ \hat{\xi}_1^\alpha \hat{\xi}_2^\beta) \hat{\xi}_2^\nu \\ \notag
&&-\hat{\xi}_1^\rho (\hat{\xi}_1^\sigma \hat{\partial}_\sigma \hat{\xi}_2^\mu - \hat{\xi}_2^\sigma \hat{\partial}_\sigma \hat{\xi}_1^\mu + \breve{f}_{\! \alpha \beta}^{\ \mu}\ \hat{\xi}_1^\alpha \hat{\xi}_2^\beta) \hat{\xi}_2^\nu \\ \notag
&&+\hat{\xi}_1^\nu (\hat{\xi}_1^\sigma \hat{\partial}_\sigma \hat{\xi}_2^\mu - \hat{\xi}_2^\sigma \hat{\partial}_\sigma \hat{\xi}_1^\mu + \breve{f}_{\! \alpha \beta}^{\ \mu}\ \hat{\xi}_1^\alpha \hat{\xi}_2^\beta) \hat{\xi}_2^\rho \\ \notag \, 
&=&-2\hat{\xi}_1^\mu (\hat{\xi}_1^\sigma \hat{\partial}_\sigma \hat{\xi}_2^\nu - \hat{\xi}_2^\sigma \hat{\partial}_\sigma \hat{\xi}_1^\nu + \breve{f}_{\! \alpha \beta}^{\ \nu}\ \hat{\xi}_1^\alpha \hat{\xi}_2^\beta) \hat{\xi}_2^\rho \\ \notag
&&-2\hat{\xi}_1^\nu (\hat{\xi}_1^\sigma \hat{\partial}_\sigma \hat{\xi}_2^\rho - \hat{\xi}_2^\sigma \hat{\partial}_\sigma \hat{\xi}_1^\rho + \breve{f}_{\! \alpha \beta}^{\ \rho}\ \hat{\xi}_1^\alpha \hat{\xi}_2^\beta) \hat{\xi}_2^\mu \\ \notag 
&&-2\hat{\xi}_1^\rho (\hat{\xi}_1^\sigma \hat{\partial}_\sigma \hat{\xi}_2^\mu - \hat{\xi}_2^\sigma \hat{\partial}_\sigma \hat{\xi}_1^\mu + \breve{f}_{\! \alpha \beta}^{\ \mu}\ \hat{\xi}_1^\alpha \hat{\xi}_2^\beta) \hat{\xi}_2^\nu  \, ,
\end{eqnarray}
where we summed over all odd permutations of indices $\mu$, $\nu$, and $\rho$, and used the fact that expressions of the type $\hat{\xi}_1^\mu (\hat{\xi}_1^\sigma \hat{\partial}_\sigma \hat{\xi}_2^\nu - \hat{\xi}_2^\sigma \hat{\partial}_\sigma \hat{\xi}_1^\nu + \breve{f}_{\! \alpha \beta}^{\ \nu}\ \hat{\xi}_1^\alpha \hat{\xi}_2^\beta) \hat{\xi}_2^\rho$ are antisymmetric with respect to interchange of $\hat{\xi}_1$ and $\hat{\xi}_2$. 

In order to show that this is in fact the expression for $\breve{R}$-flux (\ref{eq:Rbreve}), we firstly write
\begin{eqnarray} \label{eq:R0-def}
\breve{\theta}^{\mu \sigma} \hat{\partial}_\sigma \breve{\theta}^{\nu \rho} + cyclic&=& \hat{\xi}_1^\mu \hat{\xi}_2^\sigma\hat{\partial}_\sigma \hat{\xi}_1^\nu \hat{\xi}_2^\rho+\hat{\xi}_1^\mu \hat{\xi}_2^\sigma\hat{\partial}_\sigma \hat{\xi}_2^\rho \hat{\xi}_1^\nu -\hat{\xi}_1^\mu \hat{\xi}_2^\sigma\hat{\partial}_\sigma \hat{\xi}_2^\nu \hat{\xi}_1^\rho-\hat{\xi}_1^\mu \hat{\xi}_2^\sigma\hat{\partial}_\sigma \hat{\xi}_1^\rho \hat{\xi}_2^\nu \\ \notag
&&+\hat{\xi}_1^\nu \hat{\xi}_2^\sigma\hat{\partial}_\sigma \hat{\xi}_1^\rho \hat{\xi}_2^\mu + \hat{\xi}_1^\nu \hat{\xi}_2^\sigma\hat{\partial}_\sigma \hat{\xi}_2^\mu \hat{\xi}_1^\rho - \hat{\xi}_1^\nu \hat{\xi}_2^\sigma\hat{\partial}_\sigma \hat{\xi}_2^\rho \hat{\xi}_1^\mu - \hat{\xi}_1^\nu \hat{\xi}_2^\sigma\hat{\partial}_\sigma \hat{\xi}_1^\mu \hat{\xi}_2^\rho \\ \notag
&&+\hat{\xi}_1^\rho \hat{\xi}_2^\sigma\hat{\partial}_\sigma \hat{\xi}_1^\mu \hat{\xi}_2^\nu + \hat{\xi}_1^\rho \hat{\xi}_2^\sigma\hat{\partial}_\sigma \hat{\xi}_2^\nu \hat{\xi}_1^\mu - \hat{\xi}_1^\rho \hat{\xi}_2^\sigma\hat{\partial}_\sigma \hat{\xi}_2^\mu \hat{\xi}_1^\nu - \hat{\xi}_1^\rho \hat{\xi}_2^\sigma\hat{\partial}_\sigma \hat{\xi}_1^\nu \hat{\xi}_2^\mu \\ \notag
&&-\hat{\xi}_2^\mu \hat{\xi}_1^\sigma\hat{\partial}_\sigma \hat{\xi}_1^\nu \hat{\xi}_2^\rho-\hat{\xi}_2^\mu \hat{\xi}_1^\sigma\hat{\partial}_\sigma \hat{\xi}_2^\rho \hat{\xi}_1^\nu +\hat{\xi}_2^\mu \hat{\xi}_1^\sigma\hat{\partial}_\sigma \hat{\xi}_2^\nu \hat{\xi}_1^\rho+\hat{\xi}_2^\mu \hat{\xi}_1^\sigma\hat{\partial}_\sigma \hat{\xi}_1^\rho \hat{\xi}_2^\nu  \\ \notag
&&-\hat{\xi}_2^\nu \hat{\xi}_1^\sigma\hat{\partial}_\sigma \hat{\xi}_1^\rho \hat{\xi}_2^\mu - \hat{\xi}_2^\nu \hat{\xi}_1^\sigma\hat{\partial}_\sigma \hat{\xi}_2^\mu \hat{\xi}_1^\rho + \hat{\xi}_2^\nu \hat{\xi}_1^\sigma\hat{\partial}_\sigma \hat{\xi}_2^\rho \hat{\xi}_1^\mu + \hat{\xi}_2^\nu \hat{\xi}_1^\sigma\hat{\partial}_\sigma \hat{\xi}_1^\mu \hat{\xi}_2^\rho \\ \notag
&&-\hat{\xi}_2^\rho \hat{\xi}_1^\sigma\hat{\partial}_\sigma \hat{\xi}_1^\mu \hat{\xi}_2^\nu - \hat{\xi}_2^\rho \hat{\xi}_1^\sigma\hat{\partial}_\sigma \hat{\xi}_2^\nu \hat{\xi}_1^\mu + \hat{\xi}_2^\rho \hat{\xi}_1^\sigma\hat{\partial}_\sigma \hat{\xi}_2^\mu \hat{\xi}_1^\nu + \hat{\xi}_2^\rho \hat{\xi}_1^\sigma\hat{\partial}_\sigma \hat{\xi}_1^\nu \hat{\xi}_2^\mu \\ \notag
&=& \hat{\xi}_1^\mu (\hat{\xi}_2^\sigma \hat{\partial}_\sigma \hat{\xi}_1^\nu - \hat{\xi}_1^\sigma \hat{\partial}_\sigma \hat{\xi}_2^\nu )\hat{\xi}_2^\rho -  \hat{\xi}_1^\mu (\hat{\xi}_2^\sigma \hat{\partial}_\sigma \hat{\xi}_1^\rho - \hat{\xi}_1^\sigma \hat{\partial}_\sigma \hat{\xi}_2^\rho )\hat{\xi}_2^\nu \\ \notag
&&+ \hat{\xi}_1^\nu (\hat{\xi}_2^\sigma \hat{\partial}_\sigma \hat{\xi}_1^\rho - \hat{\xi}_1^\sigma \hat{\partial}_\sigma \hat{\xi}_2^\rho )\hat{\xi}_2^\mu -  \hat{\xi}_1^\nu (\hat{\xi}_2^\sigma \hat{\partial}_\sigma \hat{\xi}_1^\mu - \hat{\xi}_1^\sigma \hat{\partial}_\sigma \hat{\xi}_2^\mu )\hat{\xi}_2^\rho \\ \notag
&&+ \hat{\xi}_1^\rho (\hat{\xi}_2^\sigma \hat{\partial}_\sigma \hat{\xi}_1^\mu - \hat{\xi}_1^\sigma \hat{\partial}_\sigma \hat{\xi}_2^\mu )\hat{\xi}_2^\nu -  \hat{\xi}_1^\rho (\hat{\xi}_2^\sigma \hat{\partial}_\sigma \hat{\xi}_1^\nu - \hat{\xi}_1^\sigma \hat{\partial}_\sigma \hat{\xi}_2^\nu )\hat{\xi}_2^\mu \\ \notag
&=&-2\hat{\xi}_1^\mu (\hat{\xi}_1^\sigma \hat{\partial}_\sigma \hat{\xi}_2^\nu - \hat{\xi}_2^\sigma \hat{\partial}_\sigma \hat{\xi}_1^\nu) \hat{\xi}_2^\rho -2\hat{\xi}_1^\nu (\hat{\xi}_1^\sigma \hat{\partial}_\sigma \hat{\xi}_2^\rho - \hat{\xi}_2^\sigma \hat{\partial}_\sigma \hat{\xi}_1^\rho) \hat{\xi}_2^\mu \\ \notag 
&&-2\hat{\xi}_1^\rho (\hat{\xi}_1^\sigma \hat{\partial}_\sigma \hat{\xi}_2^\mu - \hat{\xi}_2^\sigma \hat{\partial}_\sigma \hat{\xi}_1^\mu) \hat{\xi}_2^\nu \, .
\end{eqnarray}
Secondarily, we have
\begin{eqnarray} \label{eq:Rf-def}
 - \breve{\theta}^{\mu \alpha} \breve{\theta}^{\rho \beta} \breve{f}_{\!\alpha \beta}^{\ \nu} + cyclic&=& -(\hat{\xi}_1^\mu \hat{\xi}_2^\alpha- \hat{\xi}_2^\mu \hat{\xi}_1^\alpha) (\hat{\xi}_1^\rho \hat{\xi}_2^\beta- \hat{\xi}_2^\rho \hat{\xi}_1^\beta)\breve{f}_{\!\alpha \beta}^{\ \nu}  + cyclic \\ \notag
&=&-\hat{\xi}_1^\mu \hat{\xi}_1^\rho \breve{f}_{\!\alpha \beta}^{\ \nu}\ \hat{\xi}_2^\alpha \hat{\xi}_2^\beta  + \hat{\xi}_1^\mu \hat{\xi}_2^\rho\breve{f}_{\!\alpha \beta}^{\ \nu}\ \hat{\xi}_2^\alpha \hat{\xi}_1^\beta\\ \notag
&& +\hat{\xi}_2^\mu \hat{\xi}_1^\rho\breve{f}_{\!\alpha \beta}^{\ \nu}\ \hat{\xi}_1^\alpha \hat{\xi}_2^\beta  -\hat{\xi}_2^\mu \hat{\xi}_2^\rho \breve{f}_{\!\alpha \beta}^{\ \nu}\ \hat{\xi}_1^\alpha \hat{\xi}_1^\beta + cyclic\\ \notag
&=&-2 \hat{\xi}_1^\mu \hat{\xi}_2^\rho\breve{f}_{\!\alpha \beta}^{\ \nu}\ \hat{\xi}_1^\alpha \hat{\xi}_2^\beta -2 \hat{\xi}_1^\nu \hat{\xi}_2^\mu \breve{f}_{\!\alpha \beta}^{\ \rho}\ \hat{\xi}_1^\alpha \hat{\xi}_2^\beta  - 2 \hat{\xi}_1^\rho \hat{\xi}_2^\nu \breve{f}_{\!\alpha \beta}^{\ \mu}\ \hat{\xi}_1^\alpha \hat{\xi}_2^\beta  \, ,
\end{eqnarray}
where we used the antisymmetric properties of the $\breve{f}$-flux, such as 
\begin{equation} \label{eq:R2-sim}
\breve{f}_{\!\alpha \beta}^{\ \nu}\ \hat{\xi}_1^\alpha \hat{\xi}_1^\beta = 0 \, , \quad \breve{f}_{\!\alpha \beta}^{\ \nu}\ \hat{\xi}_1^\alpha \hat{\xi}_2^\beta = -\breve{f}_{\!\alpha \beta}^{\ \nu}\ \hat{\xi}_2^\alpha \hat{\xi}_1^\beta \, .
\end{equation}
Substituting (\ref{eq:R0-def}) and (\ref{eq:Rf-def}) into (\ref{eq:Rbreve}), we obtain exactly (\ref{eq:SNB-v}). As such, we showed that (\ref{eq:breveR-thth}) stands for any simple bi-vector $\breve{\theta}$. 

\subsection{Composite bi-vector}

Let us assume that for some number $n$, the twisted Schouten-Nijenhuis bracket of a bi-vector that is composite of $n$ simple bi-vectors with itself is given by the $\breve{R}$-flux formula. If we can prove that then for $n+1$ the relation holds, according to the principle of mathematical induction this is sufficient to prove for all $n$. Let
\begin{equation}
\breve{\theta}_n = \sum_{i=1}^{n} \hat{\xi}_{i1} \wedge \hat{\xi}_{i2} \, , \qquad \breve{\theta}_{n}^{\mu \nu} = \sum_{i=1}^{n} (\hat{\xi}_{i1}^\mu \hat{\xi}_{i2}^\nu-\hat{\xi}_{i1}^\nu \hat{\xi}_{i2}^\mu) \, , 
\end{equation}
where $\hat{\xi}_{i}$ and $\hat{\zeta}_{i}$ are vectors. Assume that for that number $n$ we have
\begin{equation} \label{eq:thn-thn}
\frac{1}{2}[\breve{\theta}_n, \breve{\theta}_n]_{\hat{S}} = \breve{\theta}_n^{\mu \alpha} \hat{\partial}_\alpha \breve{\theta}_n^{\nu \rho} - \breve{\theta}_n^{\mu \alpha} \breve{\theta}_n^{\rho \beta} \breve{f}_{\!\alpha \beta}^{\ \nu} + cyclic \, .
\end{equation}
Now it is necessary to prove that for some 
\begin{equation}
\breve{\theta}_{n+1} = \breve{\theta}_n + \breve{\beta} \, , \qquad \breve{\beta}=\hat{\eta}_1 \wedge \hat{\eta}_2 \, ,  
\end{equation}
the relation (\ref{eq:thn-thn}) still holds.

We have
\begin{eqnarray} \label{eq:SNB-comp}
[\breve{\theta}_{n+1}, \breve{\theta}_{n+1}]_{\hat{S}} = [\breve{\theta}_n, \breve{\theta}_n]_{\hat{S}} + [\breve{\theta}_n, \breve{\beta}]_{\hat{S}} +[\breve{\beta}, \breve{\theta}_n]_{\hat{S}} +[\breve{\beta}, \breve{\beta}]_{\hat{S}} \, .
\end{eqnarray}
Using the graded Leibniz identity (\ref{eq:SNB-grad1}), we obtain
\begin{eqnarray} \label{eq:SNB-comp3}\notag
[\breve{\theta}_n, \breve{\beta}]_{\hat{S}} &=&\sum_{i=1}^{n}   [ \hat{\xi}_{i1} \wedge  \hat{\xi}_{i2}, \hat{\eta}_1 \wedge \hat{\eta}_2 ]_{\hat{S}} = \sum_{i=1}^{n}  \Big(  [ \hat{\xi}_{i1} \wedge  \hat{\xi}_{i2}, \hat{\eta}_1 ]_{\hat{S}} \wedge \hat{\eta}_2  - \hat{\eta}_1 \wedge [ \hat{\xi}_{i1} \wedge  \hat{\xi}_{i2}, \hat{\eta}_2  ]_{\hat{S}} \Big) \\ 
&=&\sum_{i=1}^{n}  \Big( -[\hat{\eta}_1, \hat{\xi}_{i1}]_{\hat{L}} \wedge \hat{\xi}_{i2} \wedge \hat{\eta}_2 -\hat{\xi}_{i1} \wedge [\hat{\eta}_1, \hat{\xi}_{i2}]_{\hat{L}} \wedge \hat{\eta}_2 \\ \notag
&&\qquad+\hat{\eta}_1 \wedge [\hat{\eta}_2, \hat{\xi}_{i1}]_{\hat{L}} \wedge \hat{\xi}_{i2} + \hat{\eta}_1 \wedge \hat{\xi}_{i1} \wedge [\hat{\eta}_2, \hat{\xi}_{i2}]_{\hat{L}} \Big) \, ,
\end{eqnarray}
%+ cyclic 
Furthermore, with the help of (\ref{eq:hatLie}), we firstly note that
\begin{eqnarray} \label{eq:eta-xi-1}
-\Big( [\hat{\eta}_1, \hat{\xi}_{i1}]_{\hat{L}} \wedge \hat{\xi}_{i2} \wedge \hat{\eta}_2 \Big)^{\mu \nu \rho}&=& (\hat{\xi}_{i1}^\sigma \hat{\partial}_\sigma \hat{\eta}_1^\mu-\hat{\eta}_1^\sigma \hat{\partial}_\sigma \hat{\xi}_{i1}^\mu-\breve{f}_{\! \alpha \beta}^{\ \mu}\ \hat{\eta}_1^\alpha \hat{\xi}_{i1}^\beta) \hat{\xi}_{i2}^\nu \hat{\eta}_2^\rho \\ \notag 
&&-(\hat{\xi}_{i1}^\sigma \hat{\partial}_\sigma \hat{\eta}_1^\mu-\hat{\eta}_1^\sigma \hat{\partial}_\sigma \hat{\xi}_{i1}^\mu-\breve{f}_{\! \alpha \beta}^{\ \mu}\ \hat{\eta}_1^\alpha \hat{\xi}_{i1}^\beta) \hat{\xi}_{i2}^\rho \hat{\eta}_2^\nu + cyclic\\ \notag
&=& (\hat{\xi}_{i1}^\sigma \hat{\partial}_\sigma \hat{\eta}_1^\rho-\hat{\eta}_1^\sigma \hat{\partial}_\sigma \hat{\xi}_{i1}^\rho+\breve{f}_{\! \alpha \beta}^{\ \rho}\ \hat{\xi}_{i1}^\alpha \hat{\eta}_1^\beta) \hat{\xi}_{i2}^\mu \hat{\eta}_2^\nu  \\ \notag
&& -(\hat{\xi}_{i1}^\sigma \hat{\partial}_\sigma \hat{\eta}_1^\nu-\hat{\eta}_1^\sigma \hat{\partial}_\sigma \hat{\xi}_{i1}^\nu+\breve{f}_{\! \alpha \beta}^{\ \nu}\ \hat{\xi}_{i1}^\alpha \hat{\eta}_1^\beta) \hat{\xi}_{i2}^\mu \hat{\eta}_2^\rho + cyclic \, ,
\end{eqnarray}
where we chose the index permutation that will be convenient for us later. Secondly, we note
\begin{eqnarray} \label{eq:eta-xi-2}
-\Big( \hat{\xi}_{i1} \wedge [\hat{\eta}_1, \hat{\xi}_{i2}]_{\hat{L}} \wedge \hat{\eta}_2  \Big)^{\mu \nu \rho} &=& (\hat{\xi}_{i2}^\sigma \hat{\partial}_\sigma \hat{\eta}_1^\nu - \hat{\eta}_1^\sigma \hat{\partial}_\sigma \hat{\xi}_{i2}^\nu + \breve{f}_{\! \alpha \beta}^{\ \nu}\ \hat{\xi}_{i2}^\alpha \hat{\eta}_1^\beta)\hat{\xi}_{i1}^\mu \hat{\eta}_2^\rho  \\ \notag
&&-(\hat{\xi}_{i2}^\sigma \hat{\partial}_\sigma \hat{\eta}_1^\rho - \hat{\eta}_1^\sigma \hat{\partial}_\sigma \hat{\xi}_{i2}^\rho + \breve{f}_{\! \alpha \beta}^{\ \rho}\ \hat{\xi}_{i2}^\alpha \hat{\eta}_1^\beta)\hat{\xi}_{i1}^\mu \hat{\eta}_2^\nu + cyclic.
\end{eqnarray}
In addition, we have
\begin{eqnarray} \label{eq:eta-xi-3}
\Big(\hat{\eta}_1 \wedge [\hat{\eta}_2, \hat{\xi}_{i1}]_{\hat{L}} \wedge \hat{\xi}_{i2}  \Big)^{\mu \nu \rho} &=& (\hat{\eta}_2^\sigma \hat{\partial}_\sigma \hat{\xi}_{i1}^\rho - \hat{\xi}_{i1}^\sigma \hat{\partial}_\sigma \hat{\eta}_2^\rho + \breve{f}_{\! \alpha \beta}^{\ \rho}\ \hat{\eta}_2^\alpha \hat{\xi}_{i1}^\beta) \hat{\xi}_{i2}^\mu \hat{\eta}_1^\nu\\ \notag
&&-(\hat{\eta}_2^\sigma \hat{\partial}_\sigma \hat{\xi}_{i1}^\nu - \hat{\xi}_{i1}^\sigma \hat{\partial}_\sigma \hat{\eta}_2^\nu + \breve{f}_{\! \alpha \beta}^{\ \nu}\ \hat{\eta}_2^\alpha \hat{\xi}_{i1}^\beta) \hat{\xi}_{i2}^\mu  \hat{\eta}_1^\rho   + cyclic \, , 
\end{eqnarray}
and
\begin{eqnarray} \label{eq:eta-xi-4}
\Big( \hat{\xi}_{i1} \wedge \hat{\eta}_1 \wedge [\hat{\xi}_{i2}, \hat{\eta}_2]_{\hat{L}} \Big)^{\mu \nu \rho} &=& (\hat{\xi}_{i2}^\sigma \hat{\partial}_\sigma \hat{\eta}_2^\rho-\hat{\eta}_2^\sigma \hat{\partial}_\sigma \hat{\xi}_{i2}^\rho+\breve{f}_{\! \alpha \beta}^{\ \rho}\ \hat{\xi}_{i2}^\alpha \hat{\eta}_2^\beta)\hat{\xi}_{i1}^\mu \hat{\eta}_1^\nu  \\ \notag 
&& -(\hat{\xi}_{i2}^\sigma \hat{\partial}_\sigma \hat{\eta}_2^\nu-\hat{\eta}_2^\sigma \hat{\partial}_\sigma \hat{\xi}_{i2}^\nu +\breve{f}_{\! \alpha \beta}^{\ \nu}\ \hat{\xi}_{i2}^\alpha \hat{\eta}_2^\beta) \hat{\xi}_{i1}^\mu \hat{\eta}_1^\rho + cyclic.
\end{eqnarray}
Lastly, using graded commutative identity (\ref{eq:SNB-grad2}), we obtain
\begin{equation} \label{eq:SNB-comp4}
[\breve{\beta}, \breve{\theta}_n]_{\hat{S}} = -(-1)^{1}[\breve{\theta}_n, \breve{\beta}]_{\hat{S}}  = [\breve{\theta}_n, \breve{\beta}]_{\hat{S}} \, .
\end{equation}
Combining (\ref{eq:eta-xi-1}), (\ref{eq:eta-xi-2}), (\ref{eq:eta-xi-3}), (\ref{eq:eta-xi-4}) and (\ref{eq:SNB-comp4}), as well as induction hypothesis (\ref{eq:thn-thn}) into (\ref{eq:SNB-comp}), we obtain
\begin{eqnarray} \label{eq:th-th-comp}
\frac{1}{2}\Big( [\breve{\theta}_{n+1}, \breve{\theta}_{n+1}]_{\hat{S}}\Big)^{\mu \nu \rho} &=& \breve{\theta}_n^{\mu \alpha} \hat{\partial}_\alpha \breve{\theta}_n^{\nu \rho} - \breve{\theta}_n^{\mu \alpha} \breve{\theta}_n^{\rho \beta} \breve{f}_{\!\alpha \beta}^{\ \nu} + \breve{\beta}^{\mu \alpha} \hat{\partial}_\alpha \breve{\beta}^{\nu \rho} - \breve{\beta}^{\mu \alpha} \breve{\beta}^{\rho \beta} \breve{f}_{\!\alpha \beta}^{\ \nu} \\ \notag
&&+\sum_{i=1}^{n} \Big( (\hat{\xi}_{i1}^\sigma \hat{\partial}_\sigma \hat{\eta}_1^\rho-\hat{\eta}_1^\sigma \hat{\partial}_\sigma \hat{\xi}_{i1}^\rho+\breve{f}_{\! \alpha \beta}^{\ \rho}\ \hat{\xi}_{i1}^\alpha \hat{\eta}_1^\beta) \hat{\xi}_{i2}^\mu \hat{\eta}_2^\nu\\ \notag
&&\qquad -(\hat{\xi}_{i1}^\sigma \hat{\partial}_\sigma \hat{\eta}_1^\nu-\hat{\eta}_1^\sigma \hat{\partial}_\sigma \hat{\xi}_{i1}^\nu+\breve{f}_{\! \alpha \beta}^{\ \nu}\ \hat{\xi}_{i1}^\alpha \hat{\eta}_1^\beta) \hat{\xi}_{i2}^\mu \hat{\eta}_2^\rho \\ \notag
&&\qquad+ (\hat{\xi}_{i2}^\sigma \hat{\partial}_\sigma \hat{\eta}_1^\nu - \hat{\eta}_1^\sigma \hat{\partial}_\sigma \hat{\xi}_{i2}^\nu + \breve{f}_{\! \alpha \beta}^{\ \nu}\ \hat{\xi}_{i2}^\alpha \hat{\eta}_1^\beta)\hat{\xi}_{i1}^\mu \hat{\eta}_2^\rho  \\ \notag
&&\qquad-(\hat{\xi}_{i2}^\sigma \hat{\partial}_\sigma \hat{\eta}_1^\rho - \hat{\eta}_1^\sigma \hat{\partial}_\sigma \hat{\xi}_{i2}^\rho + \breve{f}_{\! \alpha \beta}^{\ \rho}\ \hat{\xi}_{i2}^\alpha \hat{\eta}_1^\beta)\hat{\xi}_{i1}^\mu \hat{\eta}_2^\nu \\ \notag
&&\qquad+ (\hat{\eta}_2^\sigma \hat{\partial}_\sigma \hat{\xi}_{i1}^\rho - \hat{\xi}_{i1}^\sigma \hat{\partial}_\sigma \hat{\eta}_2^\rho + \breve{f}_{\! \alpha \beta}^{\ \rho}\ \hat{\eta}_2^\alpha \hat{\xi}_{i1}^\beta) \hat{\xi}_{i2}^\mu \hat{\eta}_1^\nu\\ \notag
&&\qquad-(\hat{\eta}_2^\sigma \hat{\partial}_\sigma \hat{\xi}_{i1}^\nu - \hat{\xi}_{i1}^\sigma \hat{\partial}_\sigma \hat{\eta}_2^\nu + \breve{f}_{\! \alpha \beta}^{\ \nu}\ \hat{\eta}_2^\alpha \hat{\xi}_{i1}^\beta) \hat{\xi}_{i2}^\mu  \hat{\eta}_1^\rho   \\ \notag
&&\qquad + (\hat{\xi}_{i2}^\sigma \hat{\partial}_\sigma \hat{\eta}_2^\rho-\hat{\eta}_2^\sigma \hat{\partial}_\sigma \hat{\xi}_{i2}^\rho+\breve{f}_{\! \alpha \beta}^{\ \rho}\ \hat{\xi}_{i2}^\alpha \hat{\eta}_2^\beta)\hat{\xi}_{i1}^\mu \hat{\eta}_1^\nu  \\ \notag 
&& \qquad-(\hat{\xi}_{i2}^\sigma \hat{\partial}_\sigma \hat{\eta}_2^\nu-\hat{\eta}_2^\sigma \hat{\partial}_\sigma \hat{\xi}_{i2}^\nu +\breve{f}_{\! \alpha \beta}^{\ \nu}\ \hat{\xi}_{i2}^\alpha \hat{\eta}_2^\beta) \hat{\xi}_{i1}^\mu \hat{\eta}_1^\rho + cyclic.
\end{eqnarray}

On the other hand, the $\breve{R}$-flux for this composite bi-vector can be expressed as 	
\begin{eqnarray} \label{eq:R-comp}
\breve{R}^{\mu \nu \rho} &=& (\breve{\theta}_n^{\mu \alpha} + \breve{\beta}^{\mu \alpha}) \hat{\partial}_\alpha (\breve{\theta}_n^{\nu \rho}+ \breve{\beta}^{\nu \rho})-(\breve{\theta}_n^{\mu \alpha} + \breve{\beta}^{\mu \alpha})(\breve{\theta}_n^{\rho \beta} + \breve{\beta}^{\rho \beta})\breve{f}_{\!\alpha \beta}^{\ \nu}  + cyclic \\ \notag
&=&\breve{\theta}_n^{\mu \alpha} \hat{\partial}_\alpha \breve{\theta}_n^{\nu \rho} - \breve{\theta}_n^{\mu \alpha} \breve{\theta}_n^{\rho \beta} \breve{f}_{\!\alpha \beta}^{\ \nu} + \breve{\beta}^{\mu \alpha} \hat{\partial}_\alpha \breve{\beta}^{\nu \rho} - \breve{\beta}^{\mu \alpha} \breve{\beta}^{\rho \beta} \breve{f}_{\!\alpha \beta}^{\ \nu} \\ \notag
&&+\breve{\beta}^{\mu \alpha} \hat{\partial}_\alpha \breve{\theta}_n^{\nu \rho} + \breve{\theta}_n^{\mu \alpha} \hat{\partial}_\alpha \breve{\beta}^{\nu \rho}-\breve{\theta}_n^{\mu \alpha} \breve{\beta}^{\rho \beta}\breve{f}_{\!\alpha \beta}^{\ \nu}  -\breve{\beta}^{\mu \alpha}  \breve{\theta}_n^{\rho \beta}\breve{f}_{\!\alpha \beta}^{\ \nu} + cyclic
\end{eqnarray}
We have
\begin{eqnarray} \label{eq:R12-comp1}
\breve{\theta}_n^{\mu \sigma}\hat{\partial}_\sigma \breve{\beta}^{\nu \rho}+cyclic &=&\sum_{i=1}^{n}  (\hat{\xi}_{i1}^\mu \hat{\xi}_{i2}^\sigma -\hat{\xi}_{i2}^\mu \hat{\xi}_{i1}^\sigma)\hat{\partial}_\sigma (\hat{\eta}_1^\nu \hat{\eta}_2^\rho - \hat{\eta}_2^\nu \hat{\eta}_1^\rho) + cyclic \\ \notag
&=& \sum_{i=1}^{n} \Big( \hat{\xi}_{i1}^\mu \hat{\xi}_{i2}^\sigma (\hat{\partial}_\sigma \hat{\eta}_1^\nu \hat{\eta}_2^\rho + \hat{\partial}_\sigma \hat{\eta}_2^\rho \hat{\eta}_1^\nu - \hat{\partial}_\sigma \hat{\eta}_2^\nu \hat{\eta}_1^\rho -\hat{\partial}_\sigma \hat{\eta}_1^\rho \hat{\eta}_2^\nu) \\ \notag
&& -\hat{\xi}_{i2}^\mu \hat{\xi}_{i1}^\sigma (\hat{\partial}_\sigma \hat{\eta}_1^\nu \hat{\eta}_2^\rho +\hat{\partial}_\sigma \hat{\eta}_2^\rho \hat{\eta}_1^\nu -\hat{\partial}_\sigma \hat{\eta}_2^\nu \hat{\eta}_1^\rho -\hat{\partial}_\sigma \hat{\eta}_1^\rho \hat{\eta}_2^\nu )+ cyclic \Big)
\end{eqnarray}
Similarly, we have
\begin{eqnarray}  \label{eq:R12-comp2}
\breve{\beta}^{\mu \sigma}\hat{\partial}_\sigma \breve{\theta}_n^{\nu \rho}+cyclic &=& \sum_{i=1}^{n} (\hat{\eta}_1^\mu \hat{\eta}_2^\sigma -\hat{\eta}_2^\mu \hat{\eta}_1^\sigma)\hat{\partial}_\sigma (\hat{\xi}_{i1}^\nu \hat{\xi}_{i2}^\rho - \hat{\xi}_{i2}^\nu \hat{\xi}_{i1}^\rho) + cyclic \\ \notag
&=& \sum_{i=1}^{n} \Big(\hat{\eta}_1^\mu \hat{\eta}_2^\sigma (\hat{\partial}_\sigma \hat{\xi}_{i1}^\nu \hat{\xi}_{i2}^\rho + \hat{\partial}_\sigma \hat{\xi}_{i2}^\rho \hat{\xi}_{i1}^\nu - \hat{\partial}_\sigma \hat{\xi}_{i2}^\nu \hat{\xi}_{i1}^\rho-\hat{\partial}_\sigma \hat{\xi}_{i1}^\rho \hat{\xi}_{i2}^\nu)\\ \notag
&&-\hat{\eta}_2^\mu \hat{\eta}_1^\sigma (\hat{\partial}_\sigma \hat{\xi}_{i1}^\nu \hat{\xi}_{i2}^\rho +\hat{\partial}_\sigma \hat{\xi}_{i2}^\rho \hat{\xi}_{i1}^\nu -\hat{\partial}_\sigma \hat{\xi}_{i2}^\nu \hat{\xi}_{i1}^\rho -\hat{\partial}_\sigma \hat{\xi}_{i1}^\rho \hat{\xi}_{i2}^\nu) + cyclic \Big) \\ \notag
&=& \sum_{i=1}^{n} \Big( \hat{\eta}_1^\nu \hat{\eta}_2^\sigma \hat{\partial}_\sigma \hat{\xi}_{i1}^\rho \hat{\xi}_{i2}^\mu+\hat{\eta}_1^\rho \hat{\eta}_2^\sigma \hat{\partial}_\sigma \hat{\xi}_{i2}^\nu \hat{\xi}_{i1}^\mu - \hat{\eta}_1^\nu \hat{\eta}_2^\sigma \hat{\partial}_\sigma \hat{\xi}_{i2}^\rho \hat{\xi}_{i1}^\mu \\ \notag
&& -\hat{\eta}_1^\rho \hat{\eta}_2^\sigma \hat{\partial}_\sigma \hat{\xi}_{i1}^\nu \hat{\xi}_{i2}^\mu-\hat{\eta}_2^\nu \hat{\eta}_1^\sigma \hat{\partial}_\sigma \hat{\xi}_{i1}^\rho \hat{\xi}_{i2}^\mu - \hat{\eta}_2^\rho \hat{\eta}_1^\sigma \hat{\partial}_\sigma \hat{\xi}_{i2}^\nu \hat{\xi}_{i1}^\mu \\ \notag
&&+ \hat{\eta}_2^\nu \hat{\eta}_1^\sigma \hat{\partial}_\sigma \hat{\xi}_{i2}^\rho \hat{\xi}_{i1}^\mu+\hat{\eta}_2^\rho \hat{\eta}_1^\sigma \hat{\partial}_\sigma \hat{\xi}_{i1}^\nu \hat{\xi}_{i2}^\mu+ cyclic \Big) \\ \notag
&=& \sum_{i=1}^{n} \Big[\hat{\xi}_{i2}^\mu \Big(\hat{\eta}_2^\sigma (\hat{\eta}_1^\nu \hat{\partial}_\sigma \hat{\xi}_{i1}^\rho -\hat{\eta}_1^\rho \hat{\partial}_\sigma \hat{\xi}_{i1}^\nu)-\hat{\eta}_1^\sigma (\hat{\eta}_2^\nu \hat{\partial}_\sigma \hat{\xi}_{i1}^\rho -\hat{\eta}_2^\rho \hat{\partial}_\sigma \hat{\xi}_{i1}^\nu) \Big) \\ \notag
&& + \hat{\xi}_{i1}^\mu \Big(\hat{\eta}_2^\sigma (\hat{\eta}_1^\rho \hat{\partial}_\sigma \hat{\xi}_{i2}^\nu - \hat{\eta}_1^\nu \hat{\partial}_\sigma \hat{\xi}_{i2}^\rho)- \hat{\eta}_1^\sigma (\hat{\eta}_2^\rho \hat{\partial}_\sigma \hat{\xi}_{i2}^\nu - \hat{\eta}_2^\nu \hat{\partial}_\sigma \hat{\xi}_{i2}^\rho) \Big) \Big] + cyclic\, ,
\end{eqnarray}
where we firstly manipulated with cyclic indices and then just rewrote the expression for future convenience. Lastly, we have
\begin{eqnarray} \label{eq:R12-comp3}
-\breve{\theta}_n^{\mu \alpha} \breve{\beta}^{\rho \beta}\breve{f}_{\!\alpha \beta}^{\ \nu} + cyclic &=& -\sum_{i=1}^{n} (\hat{\xi}_{i1}^\mu \hat{\xi}_{i2}^\alpha - \hat{\xi}_{i1}^\alpha \hat{\xi}_{i2}^\mu)(\hat{\eta}_1^\rho \hat{\eta}_2^\beta -\hat{\eta}_1^\beta \hat{\eta}_2^\rho)\breve{f}_{\alpha \beta}^{\ \nu} + cyclic \, , 
\end{eqnarray}
and
\begin{eqnarray} \label{eq:R12-comp4}
-\breve{\beta}^{\mu \alpha} \breve{\theta}_n^{\rho \beta}\breve{f}_{\!\alpha \beta}^{\ \nu} + cyclic &=& -\sum_{i=1}^{n} (\hat{\eta}_1^\mu \hat{\eta}_2^\alpha - \hat{\eta}_1^\alpha \hat{\eta}_2^\mu)(\hat{\xi}_{i1}^\rho \hat{\xi}_{i2}^\beta -\hat{\xi}_{i1}^\beta \hat{\xi}_{i2}^\rho)\breve{f}_{\alpha \beta}^{\ \nu} + cyclic \\ \notag
&=&-\sum_{i=1}^{n} (\hat{\eta}_1^\nu \hat{\eta}_2^\alpha - \hat{\eta}_1^\alpha \hat{\eta}_2^\nu)(\hat{\xi}_{i1}^\mu \hat{\xi}_{i2}^\beta -\hat{\xi}_{i1}^\beta \hat{\xi}_{i2}^\mu)\breve{f}_{\alpha \beta}^{\ \rho} + cyclic  \, .
\end{eqnarray}
Substituting relations (\ref{eq:R12-comp1})-(\ref{eq:R12-comp4}) into (\ref{eq:R-comp}), one obtains exactly the relation (\ref{eq:th-th-comp}). Therefore, we have demonstrated via principle of mathematical induction, that the twisted Schouten-Nijenhuis bracket of the arbitrary bi-vector with itself produces exactly the $\breve{R}$-flux.

\end{document}